\documentclass[useAMS,usenatbib,twocolumn]{mnras}
\usepackage{times}
\usepackage{graphicx}
\usepackage{lscape}
\usepackage{ amssymb }
\def\etal{{\it et al.}}

\title[Limits on FRBs and transients using the MWA]{Limits on Fast Radio Bursts and other transient sources at 182 MHz using the Murchison Widefield Array}
\author[A. Rowlinson \etal]{A. Rowlinson$^{1,2,3,4}$\thanks{E-mail:b.a.rowlinson@uva.nl},
M.~E. Bell$^{3,4}$,
T. Murphy$^{4,5}$,
C.~M. Trott$^{4,6}$,
N. Hurley-Walker$^{6}$, \and 
S. Johnston$^{3}$,
S.~J. Tingay$^{4,6}$,
D.~L. Kaplan$^{7}$,
D. Carbone$^{1}$,
P.~J. Hancock$^{4,6}$,
L. Feng$^{8}$, \and
A.~R. Offringa$^{2}$,
G.~Bernardi$^{9,10,11}$,
J.~D.~Bowman$^{12}$, 
F.~Briggs$^{13}$,
R.~J.~Cappallo$^{14}$, \and
A.~A.~Deshpande$^{15}$, 
B.~M.~Gaensler$^{4,5,16}$,
L.~J.~Greenhill$^{11}$,
B.~J.~Hazelton$^{17}$, \and
M.~Johnston-Hollitt$^{18}$,
C.~J.~Lonsdale$^{14}$, 
S.~R.~McWhirter$^{14}$,
D.~A.~Mitchell$^{3,4}$, 
M.~F.~Morales$^{17}$, \and 
E.~Morgan$^{7}$, 
D.~Oberoi$^{19}$, 
S.~M.~Ord$^{4,6}$,
T.~Prabu$^{15}$, 
N.~Udaya~Shankar$^{15}$, 
K.~S.~Srivani$^{15}$, \and 
R.~Subrahmanyan$^{4,15}$, 
R.~B.~Wayth$^{4,6}$, 
R.~L.~Webster$^{4,20}$, 
A.~Williams$^{6}$, 
C.~L.~Williams$^{7}$ \\
$^{1}$ Anton Pannekoek Institute, University of Amsterdam, Postbus 94249, 1090 GE, Amsterdam, The Netherlands\\
$^{2}$ Netherlands Institute for Radio Astronomy (ASTRON), PO Box 2, 7990 AA Dwingeloo, The Netherlands \\
$^{3}$ CSIRO Astronomy and Space Science, PO Box 76, Epping, NSW 1710, Australia\\
$^{4}$ ARC Centre of Excellence for All-sky Astrophysics (CAASTRO), Australia\\
$^{5}$ Sydney Institute for Astronomy, School of Physics, The University of Sydney, NSW 2006, Australia\\
$^{6}$ International Centre for Radio Astronomy Research (ICRAR), Curtin University, Bently, WA 6102, Australia\\
$^{7}$ Kavli Institute for Astrophysics and Space Research, Massachusetts Institute of Technology, Cambridge, MA 02139, USA\\
$^{8}$ Department of Physics, University of Wisconsin-Milwaukee, Milwaukee, WI 53201, USA\\
$^{9}$ SKA SA, 3rd Floor, The Park, Park Road, Pinelands, 7405, South Africa \\
$^{10}$ Department of Physics and Electronics, Rhodes University, PO Box 94, Grahamstown 6140, South Africa\\
$^{11}$ Harvard-Smithsonian Center for Astrophysics, Cambridge, MA 02138, USA\\
$^{12}$ School of Earth and Space Exploration, Arizona State University, Tempe, AZ 85287, USA\\
$^{13}$ Research School of Astronomy and Astrophysics, Australian National University, Canberra, ACT 2611, Australia\\
$^{14}$ MIT Haystack Observatory, Westford, MA 01886, USA\\
$^{15}$ Raman Research Institute, Bangalore 560080, India\\
$^{16}$ Dunlap Institute for Astronomy and Astrophysics, University of Toronto, ON, M5S 3H4, Canada\\
$^{17}$ Department of Physics, University of Washington, Seattle, WA 98195, USA\\
$^{18}$ School of Chemical \& Physical Sciences, Victoria University of Wellington, Wellington 6140, New Zealand\\
$^{19}$ National Centre for Radio Astrophysics, Tata Institute for Fundamental Research, Pune 411007, India\\
$^{20}$ School of Physics, The University of Melbourne, Parkville, VIC 3010, Australia
}

\begin{document}

\pagerange{\pageref{firstpage}--\pageref{lastpage}} \pubyear{000}
\maketitle            

\label{firstpage}

\begin{abstract}

We present a survey for transient and variable sources, on timescales from 28 seconds to $\sim$1 year, using the Murchison Widefield Array (MWA) at 182 MHz. Down to a detection threshold of 0.285 Jy, no transient candidates were identified, making this the most constraining low-frequency survey to date and placing a limit on the surface density of transients of $<4.1 \times 10^{-7}$ deg$^{-2}$ for the shortest timescale considered.
At these frequencies, emission from Fast Radio Bursts (FRBs) is expected to be detectable in the shortest timescale images without any corrections for interstellar or intergalactic dispersion. At an FRB limiting flux density of 7980 Jy, we find a rate of $<$82 FRBs per sky per day for dispersion measures $<$700 pc cm$^{-3}$. Assuming a cosmological population of standard candles, our rate limits are consistent with the FRB rates obtained by \cite{thornton2013} if they have a flat spectral slope.
Finally, we conduct an initial variability survey of sources in the field with flux densities $\gtrsim$0.5 Jy and identify no sources with significant variability in their lightcurves. However, we note that substantial further work is required to fully characterise both the short term and low level variability within this field.

\end{abstract}

\begin{keywords}
instrumentation: interferometers - techniques: image processing - catalogues - radio continuum: general
\end{keywords}

\section{Introduction}

Until recently, little was known about the population of transient sources at low radio frequencies due to the lack of previous dedicated, sensitive surveys. Many of the known target transient populations are synchrotron sources, hence predicted to be faint and vary on long timescales at low radio frequencies \citep[such as afterglows from gamma-ray bursts and tidal disruption events; for a recent review see][]{metzger2015}. However, there are a number of different populations of sources that are expected to emit short duration bursts of low frequency coherent radio emission and are anticipated to be detectable in short snapshot low radio frequency images \citep[e.g. giant pulses from pulsars and flares from brown dwarfs or exoplanets;][]{bastian2000, berger2001, law2011, jaeger2011, murphy2015}.

One such coherently emitting target is the population of Fast Radio Bursts \citep[FRBs;][]{lorimer2007,thornton2013}. FRBs were discovered at 1.4 GHz using high time resolution observations from the Parkes radio telescope. These sources constitute single, non-repeating, bright pulses of millisecond duration at 1.4 GHz that are highly dispersed, suggesting an extra-galactic origin. A number of theories have been proposed as the progenitors of FRBs, including both extra-galactic \citep[e.g.][]{kashiyama2013,totani2013,falcke2014,lyubarsky2014,zhang2014} and Galactic origins \citep[e.g.][]{loeb2014}. The scattering for FRBs is highly dependent upon the observing frequency and is expected to smear out the pulse to much longer durations at low radio frequencies \citep[][]{hassall2013,trott2013}. The pulse durations at low radio frequencies make them more difficult to detect using standard search methods at high time resolution. Instead, their durations are expected to be comparable to those attainable in short snapshot images. However, it is unclear what the rates of FRBs at low frequencies will be because the rates are still being constrained at higher frequencies and little is known about their spectral shape \citep[e.g.][]{keane2015,karastergiou2015}. Therefore, observations at low frequencies will aid in constraining both the rates and the spectral slopes of FRBs. By more tightly constraining the rates, some progenitor mechanisms may be ruled out, including those associated with other populations with relatively low rates \citep[such as short gamma-ray bursts;][]{zhang2014}. Additionally all FRBs to date have been detected using single dish telescopes leading to large positional uncertainties \citep[e.g. 14 arcmin;][]{thornton2013}. By detecting FRBs in short snapshot image plane data observed using a low frequency radio interferometer, their positions can be constrained to higher accuracy ($<$1 arcmin) enabling host galaxy associations and deep constraints on multi-wavelength counterparts. Additionally, an interferometer will obtain more reliable flux densities, as single dish observations are subject to flux density uncertainties as the position of the source within the primary beam is unknown. This provides better constraints on the flux density distribution of sources ($\log N$--$\log S$ distribution).

Over the past few years, the search for transient sources at low radio frequencies has intensified with the arrival of sensitive, wide-field instruments such as the Murchison Wide-field Array \citep[MWA;][]{lonsdale2009,tingay2013}, the Low Frequency Array \citep[LOFAR;][]{haarlem2013} and the Long Wavelength Array Station 1 \citep[LWA1;][]{ellingson2013}. Additionally, the automated processing of very large datasets is being enabled via the long-term source monitoring capabilities of specially developed pipelines, including the LOFAR Transients Pipeline \citep[{\sc TraP};][]{swinbank2015} and the pipeline for the ASKAP Survey for Variables and Slow Transients \citep[VAST;][]{murphy2013}. Dedicated transient surveys are utilising the improvement in instrumentation and software to constrain the surface densities of transients at these low frequencies on a range of timescales and sensitivities \citep[e.g.][]{bell2014,carbone2014,obenberger2015,stewart2015}. Orders of magnitude improvement in sensitivity or search area will be required to more tightly constrain their rates. This can be attained by the next generation of radio telescopes, such as the Square Kilometre Array \citep[SKA; e.g. ][]{fender2015}. However, obtaining the required observation time may be difficult on over-subscribed instruments and transient surveys will need to utilise commensal observations. This paper uses observations from one such dataset, the MWA observations for the campaign to detect the Epoch of Re-ionisation (EoR) in which hundreds of hours of observing time are required on individual fields. This dataset can probe variability and transients on timescales ranging from seconds up to years, enabling constraints to be placed on both the long timescale incoherent emission mechanisms, e.g. synchrotron emission from Active Galactic Nuclei (AGN), short timescale coherent emission mechanisms such as FRBs and scintillation processes on a range of timescales.

This paper describes a pilot transient and variability search using 78 hours of the MWA EoR dataset, producing highly competitive transient rates. The 28 second snapshot timescale is chosen to specifically target the expected population of FRBs. This work complements \cite{tingay2015}, a search for FRBs using MWA observations imaged on a much shorter integration time (2 seconds) and conducting an image plane de-dispersion to search for FRBs. Via this method, \cite{tingay2015} are able to attain an improvement in sensitivity for FRBs in comparison to the standard processing strategies at the expense of processing speed and resolution. Whereas a standard imaging strategy, such as that utilised in this paper, enables more observations to be processed in a comparable timescale and the use of the data products for additional science such as longer duration transient and variability studies. Without de-dispersion, a dispersed FRB will be detected at a lower flux density in the short snapshot images as the original signal is averaged over both time and frequency. Therefore, these two approaches are complementary; \cite{tingay2015} increases sensitivity by sacrificing surveyed area whereas the survey conducted in this paper sacrifices sensitivity to increase the amount of surveyed area. Additionally, candidate FRBs identified in this analysis can be independently confirmed as FRBs by measuring their dispersed signal using the pipeline developed \cite{tingay2015}.

Section 2 of this paper describes the processing strategies used to make all the images and the analysis strategies implemented to conduct quality control and to search for transient sources. In Section 3, we present the limits on transients detected on a range of timescales and focus on the implications for the rates and spectral shapes of FRBs by comparison to previous studies at other frequencies. Finally, Section 4 provides an initial analysis of variability of known sources within the field.

\section{Observations and processing method}

\subsection{Dataset}

The data used in this paper are obtained from a commensally observed dataset for the transients team and the EoR team. The full dataset comprises of $>$1000 hours targeting 3 specific fields, well off the Galactic plane, centred on 2 different observing frequencies (154 and 182 MHz). Processing the full dataset is a ``big data'' scale computational challenge due to the supercomputing time required and the data volume at all of the processing stages. By targeting a subsample of the dataset, we can develop automated strategies to make the data volume more manageable and quantify the supercomputing requirements for the full dataset. In this study, we choose observations of a single target field, centred on RA: 0.00 deg, Dec: -27.00 deg  (00:00:00, $-$27:00:00; J2000), at the observing frequency of 182 MHz and at elevations of $>$75 degrees. This field is centred on the Galactic co-ordinates l: 30.636 deg, b: -78.553 deg (30:38:08.4, -78:33:10.6). The observations were conducted by taking multiple pointed observations as the field drifts through 5 different azimuth-elevation pointing directions centred on zenith. These observations were then phase centred to RA: 0.00 Dec: $-$27.00 deg (J2000), with a primary beam half width half maximum (HWHM) of 11.3 degrees.

This leads to a sample size of 3010 individual observations of 2 minute integration times, or 100 hours, in the time range 2013 August 23 -- 2014 September 14.

\subsection{Imaging strategy}

\begin{table}
\centering
\caption{The {\sc WSClean} settings used to image all the observations presented in this analysis. All other settings were the default settings.} \label{table:imgSettings} 
\begin{tabular}{|l|c|}
\hline
Setting                            & Value   \\
\hline
UV range (k$\lambda$)              & $>$0.03 \\
Maximum number of clean iterations & 20000   \\
Size of image (pixels)             & 3072    \\
Size of one pixel (arcsec)         & 54      \\
Stopping threshold for Clean (Jy beam$^{-1}$)  & 0.2     \\
Briggs weighting                   & -1      \\
\hline
\end{tabular}
\end{table}

\begin{table}
\centering
\caption{The calibrator observations used in this analysis. Observation ID 1061661200 is the principle calibration dataset, used to calibrate the rest of the dataset, and was calibrated using the observation of 3C444.} \label{table:calibrators} 
\begin{tabular}{|l|c|c|c|}
\hline
Observation ID & Time & Azimuth & Elevation \\
 & (UT) & (degrees) & (degrees) \\
\hline
1061650704 (3C444) & 2013-08-27 14:58:08 & 63.43   & 74.63 \\
1061657536         & 2013-08-27 16:52:00 & 90.00   & 76.28 \\
1061659368         & 2013-08-27 17:22:32 & 90.00   & 83.19 \\
1061661200         & 2013-08-27 17:53:04 & 0.00    & 90.00 \\
1061663032         & 2013-08-27 18:23:36 & 270.00  & 83.19 \\
1061664856         & 2013-08-27 18:54:00 & 270.00  & 76.28 \\
\hline
\end{tabular}
\end{table}

Our imaging strategy builds upon the MWA imaging pipeline developed for transient searches by \cite{bell2014}. As described in this section, we have updated and adapted this pipeline to utilise calibration and imaging tools specifically developed for the MWA, targeting a consistent flux density scale across the dataset. All data are initially processed by the MWA pre-processing tool {\sc Cotter} (Version 3.3), which includes radio frequency interference (RFI) flagging by {\sc AOFlagger} \citep[Version 2.6.1;][]{offringa2010,offringa2015}.

We conduct an initial calibration on one of the observations at zenith (observation id: 1061661200, selected at random from a sample of good observations), with an integration time of 122 seconds, from the night of 2013 August 27 by transferring a time-independent, frequency dependent calibration solution from a two minute observation of 3C444. The calibrator 3C444 was chosen because it has a good calibration model, which is based on an extrapolation of its VLSSr \citep[the Very Large Array Low-frequency Sky Survey Redux;][]{lane2014} flux density using a spectral index of -0.95. The calibrated observation of the target field is imaged using {\sc WSClean} \citep[Version 1.6][]{offringa2014} using the settings in Table \ref{table:imgSettings}. The resulting model image is primary beam corrected using MWA specific tools (Version 1.1.0) to give a model image of the field. To optimise calibration and imaging for different Az-El (azimuth-elevation) pointing directions, we created calibration images from the same night for each of the unique lower elevation pointing directions where the pointing direction is approximately equal to that of ID 1061661200\footnote{Calibration can fail for observations on different nights, with differing Az-El and RA-Dec pointing directions due to the increased complexity and uncertainties in beam models. By choosing the same night and RA-Dec pointing we can achieve a reliable calibration image for each Az-El pointing that can be applied to other nights.}. These observations, presented in Table 4, are calibrated using the 122 second image from ID 1061661200 and the phase and amplitude calibration method described below, producing a total of 5 calibration images for the full observations. This produces a consistent flux density scale across the calibration images for each of the pointing directions used in this dataset.

For each unique Az-El pointing direction on each observing night, the target RA-Dec pointing direction changes by $\sim$3 degrees as the field tracks through the Az-El pointing direction. The observations are then shifted to a common RA-Dec phase centre of 0.00,$-$27.00. This leads to the centre of the primary beam shifting with respect to the centre of the image. To enable direct comparisons between images, avoiding the lower sensitivity regions, we use a conservative 12 degree radius (approximately the primary beam HWHM) from the centre of the image for the transient and variability analysis.

We calculate the MWA primary beam for each unique observation and, using this new primary beam, un-correct the relevant model calibrator Stokes-I image (taken from Table \ref{table:calibrators}). This gives a non-primary beam corrected model image of the expected sky that the MWA has observed. For the initial calibration of the non-zenith observations in Table \ref{table:calibrators} we used the zenith image (observation id: 1061661200) and, following the creation of these calibrator images, for the imaging of the full dataset we used the calibrator image corresponding to the identical Az-El pointing direction. Using the imager, {\sc wsclean}, this model image is converted to a clean component image which can be input into the calibration tools as a sky model. With the MWA specific tools {\sc calibrate} and {\sc applysolutions}, we complete a phase and amplitude self calibration on each observation using the newly created sky model\footnote{Each observation ID is hence calibrated with a unique sky model which is based upon the calibration image.}.

We note that the field of view will likely include different isoplanatic patches within the ionosphere, which can lead to issues when conducting self calibration \citep[e.g.][]{lonsdale2005}. However, due to the compactness of the MWA, the full array observes the same isoplanatic patches \citep[regime 3 from ][]{lonsdale2005}. This leads to the apparent positions of sources varying with time, but no deformations of sources \citep{intema2009}. Due to the large number of sources spread across the sky model image, the MWA calibration strategy can account for this shift in position \citep[e.g.][]{morales2005}. Additionally, as shown by \cite{loi2015b}, under normal ionospheric conditions, the typical positional shift due to the ionosphere at 182 MHz is $\sim$10 arcsec, which is less than the resolution of the array. Therefore, the self calibration strategy outlined here is not likely to be significantly affected by viewing different isoplanatic patches, although it may to lead to small positional shifts in some of the sources.

Following calibration, we image each observation in 4 parts, corresponding to 28 second integration times, using the settings in Table \ref{table:imgSettings} and a primary beam correction using the MWA specific tool {\sc beam}.

\subsection{Initial image rejection}

\begin{figure}
\centering
\includegraphics[width=0.48\textwidth]{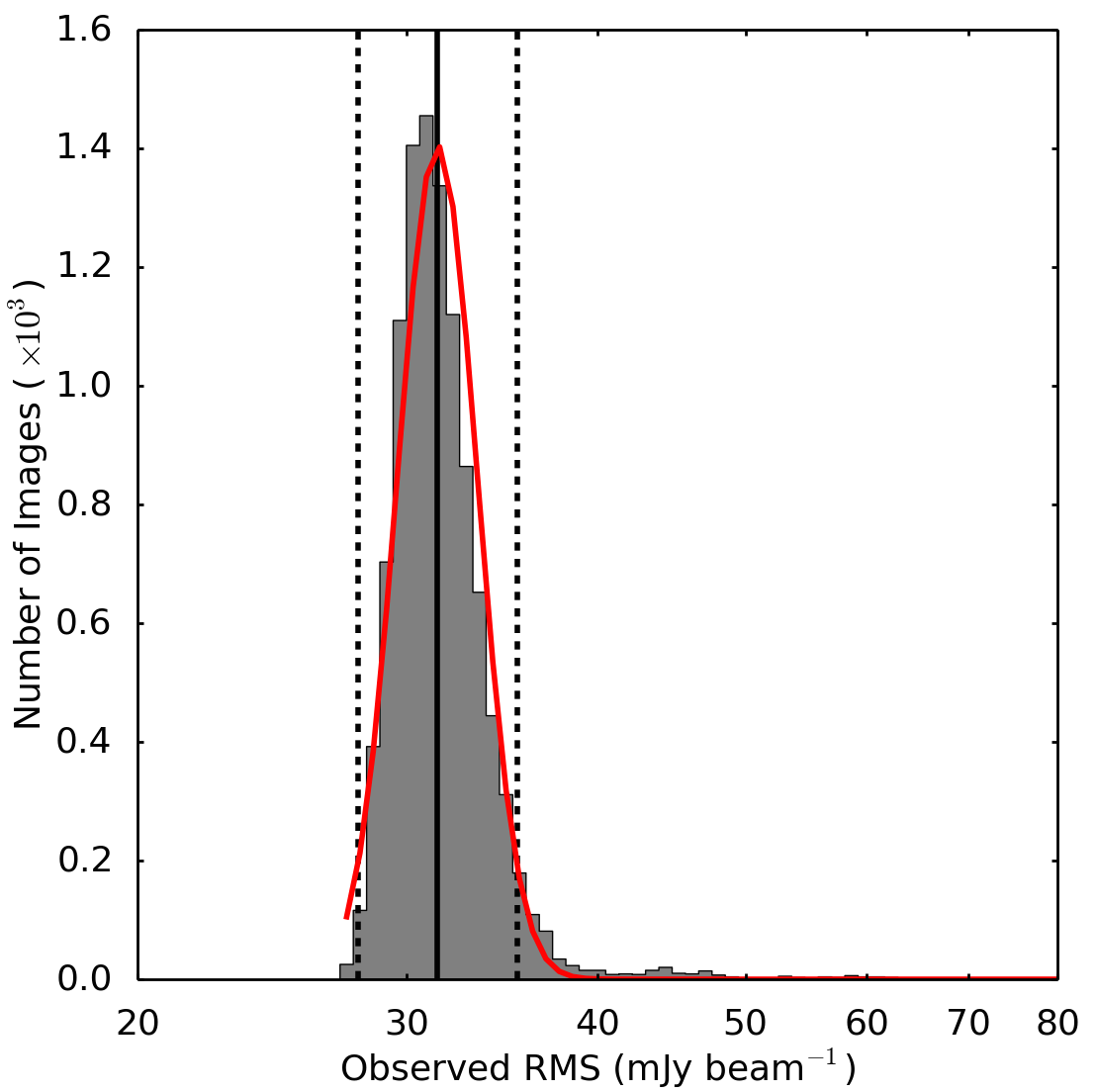}
\caption{A histogram of the RMS noise measured in the images and fitted with a Gaussian distribution (solid red line). The solid black line is the mean RMS observed, $\sim31$ mJy beam$^{-1}$, and the black dashed lines represent the 2$\sigma$ quality control rejection thresholds. $\sim$5\% of the images were rejected by this analysis.}
\label{RMShist}
\end{figure}

The imaging strategy described in Section 2.2 is on the whole very successful, leading to a large number of images with consistent properties. However, on some occasions the images will not be of sufficient quality for transient and variability searches, for a variety of reasons such as calibration errors or significant ionospheric activity \citep[e.g.][]{loi2015}, and we want to remove these images from the sample. The root mean square (RMS) noise is a powerful indicator of the quality of the image, where RMS values that deviate significantly from the expectations indicating that there are problems such as high RFI or calibration errors. We measure the mean RMS noise in the central $\frac{1}{8}$ region of each of the images using the method described in \cite{swinbank2015}, where the sources have been excluded by rejecting pixels that are 4$\sigma$ above the median RMS.

A number of images had RMS values that were highly deviant from the general population and some observations failed to image. These extremely low quality images were from the same nights, including the night of 2013 October 15, which was demonstrated to have significant ionospheric activity by \cite{loi2015}. All observations from these bad nights were removed from the sample. 

The RMS values of the remaining 10,615 images are plotted in the histogram shown in Figure \ref{RMShist} and are fitted with a Gaussian distribution\footnote{To obtain the optimal Gaussian fit to the histogram, we do not fit images with RMS $>3\sigma$ from the mean RMS}. The RMS distribution is sharply peaked with a typical RMS of $31.4^{+1.9}_{-1.8}$ mJy beam$^{-1}$. MWA becomes confusion limited in images with an integration time on the order of 2 minutes and, hence, these images do not reach the classical confusion noise in the 30 second snapshot images. Additionally, due to their short integration times, these images are dominated by sidelobe confusion which is difficult to quantify. This result is unsurprising as all images are from the identical field with the same imaging settings, therefore we can easily identify low quality images as outliers to this distribution. We reject all images with RMS values in excess of $2\sigma$ from the Gaussian distribution (corresponding to the dashed lines in Figure \ref{RMShist}) as this limit corresponds to where the observed distribution is deviating from the Gaussian distribution. Following this quality control step, 10,122 images remain in the sample used in the remainder of this paper.

\subsection{Correlated noise between observations}

\begin{figure}
\centering
\includegraphics[width=0.48\textwidth]{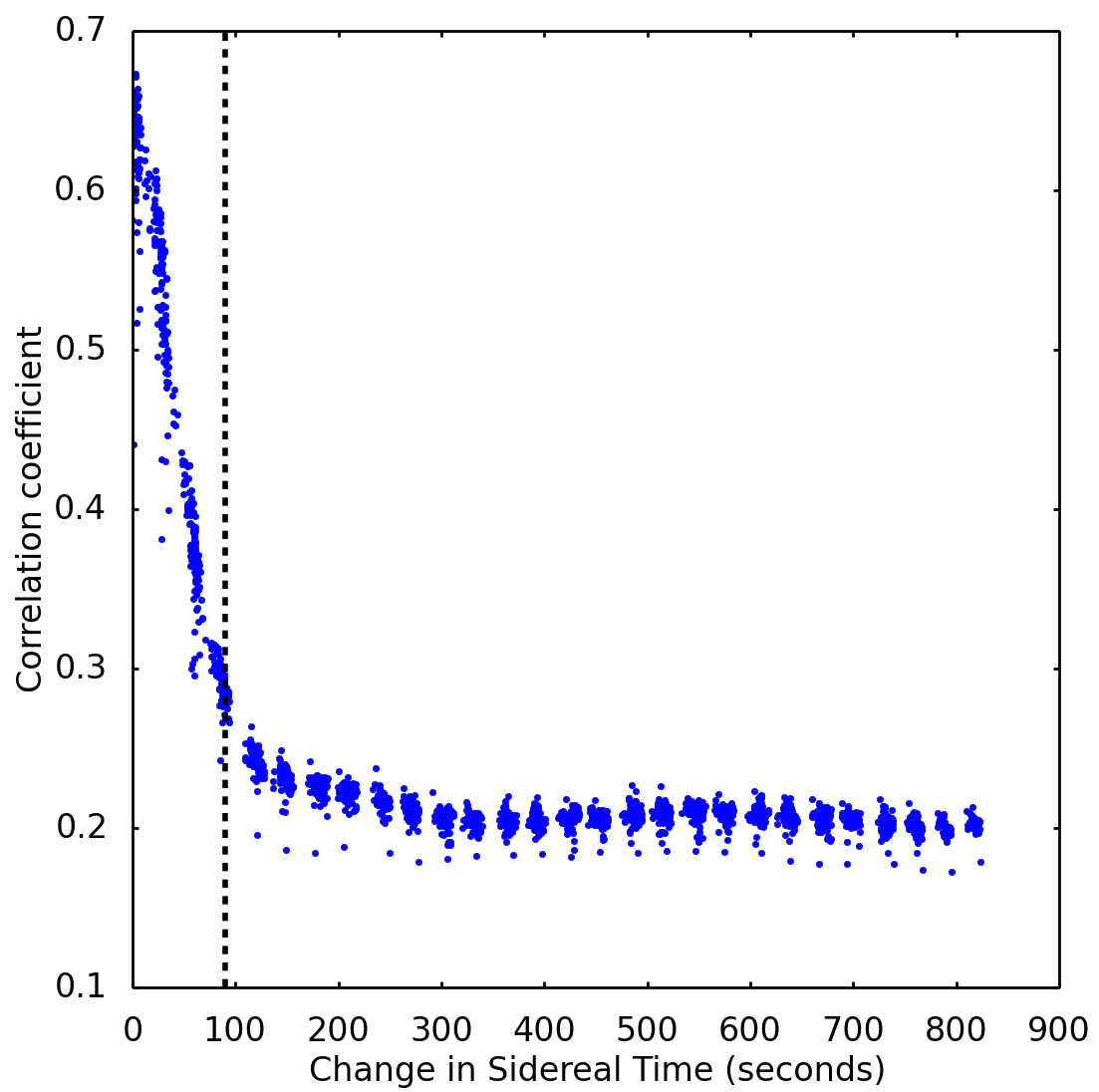}
\caption{A plot showing the correlation coefficient for pixels between the first zenith snapshot image and all other zenith snapshot images as a function of the change in local sidereal time between the two images. For images less than 90 seconds apart in local sidereal time, shown by the black dashed line, correlated noise will dominate.}
\label{correlatedNoise}
\end{figure}

\begin{figure*}
\centering
\includegraphics[width=0.95\textwidth]{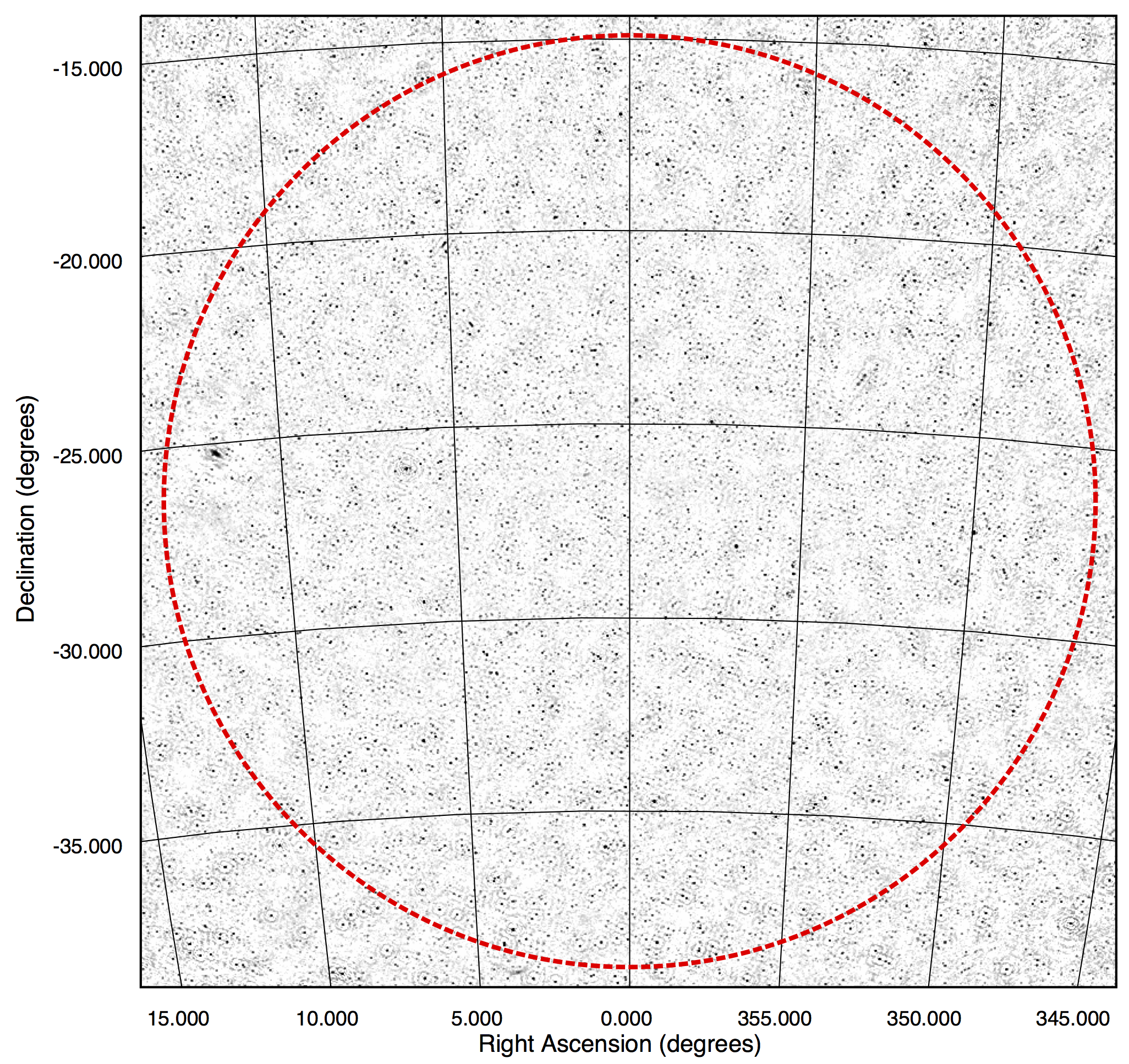}
\caption{The median image created using 10,122 images (excluding the bad images following the quality control in Section 2.4) of the field with a J2000 co-ordinate grid overlaid (in degrees), using a logarithmic colour scale. The dashed red circle shows the 12 degree radius source extraction region used in the analysis conducted in this paper.}
\label{SimTran}
\end{figure*}

The RMS noise in each of the images in the sample, characterised in Section 2.3, is made up of a number of components including the thermal noise, classical confusion noise and side-lobe confusion noise. The thermal noise component consists of a Gaussian random noise term made up of the sky noise and receiver noise, which is $\sim$2 mJy beam$^{-1}$ at 182 MHz \citep{wayth2015}. Due to the compactness of the MWA, classical confusion noise dominates the thermal noise component as multiple faint sources will be detected within the restoring beam. This component can be estimated using source counts from surveys \citep{condon1974}. The density of faint sources at 182 MHz is currently poorly known and we need to extrapolate from surveys at other frequencies. Here we calculate the classical confusion noise by extrapolating the source counts from the VLSS survey at 74 MHz \citep[][]{cohen2007} assuming a spectral index of -0.7, using
\begin{eqnarray}
\sigma_{\rm conf} = 29 \left( \frac{\theta}{1''} \right) ^{1.54} \left( \frac{\nu}{74 ~{\rm MHz}} \right) ^{-0.7} ~{\rm \mu Jy~beam^{-1}} \label{confusion}
\end{eqnarray}
where $\sigma_{\rm conf}$ is the classical confusion noise, $\theta$ is the size of the restoring beam and $\nu$ is the observing frequency. At our survey frequency of 182 MHz, the classical confusion noise is estimated to be 7 mJy. The 28 second images presented in this survey are therefore dominated by side-lobe confusion and other sources of noise. The side-lobe confusion noise is caused by imperfect deconvolution of sources either within the image, which we have minimised by using a self-calibration strategy, or in the side-lobes of the primary beam. Both of the sources of confusion noise are examples of correlated noise, where the noise for specific positions within the images between consecutive images are related to each other.

MWA images are subject to having correlated noise caused by sources drifting through the side-lobes of the primary beam (side-lobe confusion) and these images are likely to have correlated noise caused by the classical confusing sources below the detection threshold. The correlated noise component caused by side-lobe confusion will most strongly affect images that are close together in time and particularly those with similar local sidereal times, because the position of sources in the side-lobes will be essentially identical. Correlated noise can be quantified by measuring the correlation coefficient for pixels\footnote{Again removing sources by filtering out the pixels $>4\sigma$ from the median value, as utilised in the analysis in Section 2.4.} between two different images. To investigate the characteristic time-scale of this correlated noise, we focus on a single dataset centred on the zenith pointing direction and compare the first image to every other observation in that dataset. In Figure \ref{correlatedNoise}, we plot the correlation coefficient between the first image and every other image as a function of the difference between the local sidereal time for the two observations. Figure \ref{correlatedNoise} clearly shows there is significant correlated noise between two images separated by $<$90 seconds in local sidereal time. Additionally, there is a small level of correlated noise present in all the images, this is caused by a number of sources in the field that are just below the detection threshold.

The presence of correlated noise means that two observations that are close to each other in time are not  statistically independent. Therefore, when the light curve of sources are processed, data points that are close in time are not statistically independent and can bias the reduced weighted $\chi^{2}$ (which we define as $\eta$) that is typically used to identify variable sources causing issues with variability studies \citep{bell2014}. This can be corrected for in the variability statistics by reducing the number of degrees-of-freedom. Correlated noise dominates for images separated by $<$90 seconds, this corresponds to 3 snapshot images and reduces the number of degrees-of-freedom used in computing $\eta$ by a factor of order 3. As this paper focuses on a dataset monitoring the identical field and sources typically have the same number of data points, this will lead to a systematic shift in the distributions \citep{bell2014} but variable sources will still be anomalous to the distribution. In conclusion, we note that there is correlated noise in this dataset, which will be important for detailed intrinsic variability studies, but does not hinder the identification of significantly variable sources as targeted in Section 4 of this paper.

\subsection{Median Image}

A simple method to confirm the detection of transient sources is to compare the candidate to a deeper image of the same region. To obtain this deep image, we use all the 28 second images that passed the quality control strategies described in Section 2.3. As the images are already on a common co-ordinate grid, centred at an RA and Dec of 0.0,$-$27.0, we can simply average the pixel values by taking the mean or the median value of each pixel. We choose to produce the median image, as poor quality images will not bias the median image but could affect the mean image. A bright transient source that is ``on'' for a small number of images, may leave a residual source in a mean image. However, when computing the median image, we may lose some of the extended flux density as the broader regions of the point spread function (PSF) may not combine neatly. This does not significantly affect the point sources so will have a negligible affect on our analysis. 

The final median image produced, shown in Figure \ref{SimTran}, reaches an RMS noise of $\sim$8 mJy beam$^{-1}$ in the inner $\frac{1}{8}$ region, which is consistent with that expected from the combination of thermal noise and classical confusion noise (9 mJy, assuming a typical spectral slope and source counts from the VLSS at 74 MHz, see Section 2.4). Therefore, this method has significantly reduced the background noise as a number of noise components such as side-lobe confusion are averaged out due to the large number of images at different local sidereal times and pointing directions.

We identify 5548 unique sources in the median image (using the settings in Table \ref{table:SFsettings}, see Section 2.6) and we can compare this to the number of sources that are expected to be detected in this image using the VLSSr \citep{lane2014} catalogue number counts. This is the ideal survey for comparison as it is at a reasonably close observing frequency and has a comparable resolution to our images. VLSSr detected 92,964 sources in a sky area of 9.38 steradians, equivalent to $\sim$3 sources per square degree, at a sensitivity of $\sim$450 mJy. Here, we assume that the median image has a consistent RMS of 8 mJy, which is a reasonable assumption due to the observing strategy for this field and our choice to focus on the inner 12 degree radius, so much of the primary beam response has been averaged out. By assuming a spectral slope of $-0.7$, we can scale the sensitivity of the median image at 182 MHz to the frequency of the VLSSr, giving a sensitivity of 90 mJy at 74 MHz. We can then utilise the relation
\begin{eqnarray}
N \propto S^{0.9}, \label{eqn:surveyCount}
\end{eqnarray}
where $N$ is the number of sources expected in the survey and $S$ is the sensitivity of the survey. We find that we expect to detect 5810 sources in the median image, therefore we detect 95\% of the sources expected. The number of sources detected is slightly lower than expected as we have not accounted for any residual differences in the primary beam response across the image and we have assumed a single spectral slope of -0.7 can represent the full population of sources.
We also compare the number of sources detected in the median image to the 7C catalogue \citep{mcgilchrist1990} and determine that we detect an equivalent number of sources down to the completeness sensitivity of the 7C catalogue.

\subsection{The LOFAR Transients Pipeline}

\begin{table}
\centering
\caption{The {\sc TraP} source finder settings used in the transient search (Section 3). All other settings were the default settings. } \label{table:SFsettings} 
\begin{tabular}{|l|c|c|}
\hline
Setting                            & Transient Search \\
\hline
Detection Threshold                & 6$\sigma$ \\
Analysis Threshold                 & 4$\sigma$ \\
Grid size                          & 100 pixels \\
Deblend Thresholds                 & 1 \\
Gaussian fit using restoring beam & False \\
Source extraction radius           & 12 degrees ($\sim$primary beam HWHM) \\
\hline
\end{tabular}
\end{table}

To process the images, we used the LOFAR Transients Pipeline \citep[{\sc TraP}, Release 2.0;][]{swinbank2015}. In the following sections, we utilised {\sc TraP} default settings unless stated otherwise\footnote{For further details about these capabilities, refer to {\sc TraP} documentation at http://tkp.readthedocs.org/en/release2.0/.}.

The source finder used in {\sc TraP} has been optimised for transient searches \citep[][]{spreeuw2010}. Gaussians are fitted to each detected source and the fitted parameters (size and orientation) can significantly vary for a point source between consecutive images, due to the noise properties in the surrounding region, leading to artificial variability in the flux density of the source. Variable sources are expected to be point sources, so we can assume they take the shape of the restoring beam to mitigate against this problem. This strategy leads to underestimation of the flux densities of extended sources, but this is a reasonable sacrifice for variability searches as their flux densities will remain stable. A further underestimation of the flux densities can be caused by the ionosphere making point sources slightly larger than the restoring beam size; typically this effect is negligible but can become larger on nights of high ionospheric activity (these nights are typically rejected from the dataset; see Section 2.3).

However, by forcing all the sources to take the shape of the restoring beam, the source finder can fail to converge to a solution and affect the completeness of the sample, especially at low signal to noise ratios (SNRs) and in images with a small number of pixels across the restoring beam shape. For the images used in this analysis, we found the completeness was significantly reduced when fitting sources where the Gaussian shape parameters in the least-squares fit were fixed to take the shape of the restoring beam.

Therefore, to maximise both the completeness of the survey for faint transient sources and have reliable flux densities, we have used two different strategies to process the images: 

\begin{itemize}
    \item Standard: {\sc TraP} conducts a blind and unconstrained source extraction on each individual image and produces a light curve for every source detected. This setting was used for the transient search conducted in Section 3 with the parameters given in Table \ref{table:SFsettings}.
    \item Monitoring: A list of positions are given as an input to {\sc TraP}, which then fits a Gaussian using a least-squares method, with the Gaussian shape being held constant at fixed values equal to that of the restoring beam, at each position in every image to produce a reliable light curve for each source. The position is also held approximately constant during the fitting procedure, with a 10 pixel variation allowed. This setting was used for the flux density stability tests in Section 2.9 and the variability search conducted in Section 4.
\end{itemize}

\subsection{Source finder performance}
\begin{figure}
\centering
\includegraphics[width=0.48\textwidth]{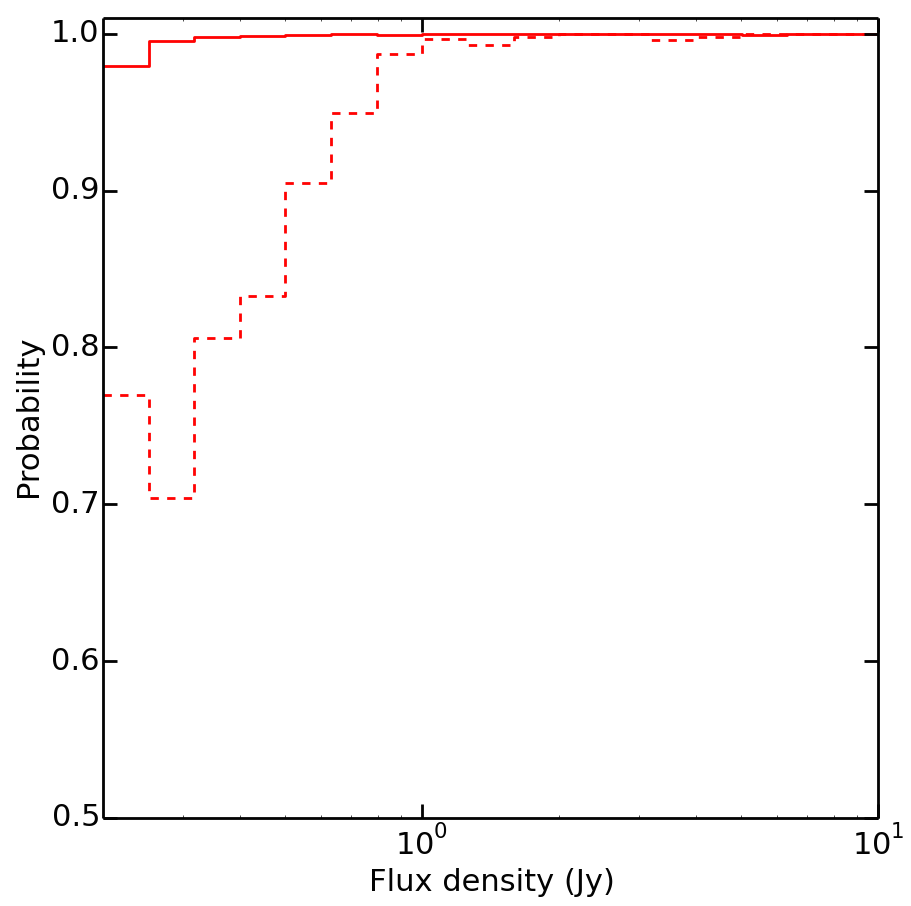}
\caption{The precision (solid line) and recall (dashed line) of the source finder settings used in this paper as a function of the flux density of the sources. The precision is roughly constant at $\sim$100\% whereas the recall drops to below 90\% for flux densities $<$0.5 Jy.}
\label{sourceFinderTest}
\end{figure}

We tested the completeness and reliability of the {\sc TraP} source finder settings used for the transient search. The source finder used in this section is the stand-alone, command-line version of the source finder used by {\sc TraP} (known as {\sc PySE}). 

We extracted sources from 250 randomly chosen images from the dataset (those passing the quality control in Section 2.3), using the settings summarised in Table \ref{table:SFsettings} and compared them to the sources detected in the significantly deeper median image using a source association radius equivalent to 10 pixels. We raised the detection threshold for the median image to 18$\sigma$ such that a source detected at 6$\sigma$ in the lowest RMS image in the sample (mean RMS - 2$\sigma$, see Figure \ref{RMShist}) would be easily detected in the median image. By raising the analysis threshold from 3$\sigma$ to 4$\sigma$, we were able to reduce the number of spuriously large ellipses fitted to noise artefacts by the source finder. For evenly spaced signal to noise ratio (SNR) bins, we counted the number of true positive (TP) detections where we correctly found the source, the false positive (FP) detections which are spurious sources and the false negative (FN) detections which are the missed sources. For each flux density bin we calculated the precision given by:
\begin{eqnarray}
{\rm Precision} = \frac{{\rm TP}}{{\rm TP}+{\rm FP}} \equiv 1 - {\rm FDR}, \label{precision}
\end{eqnarray}
where the FDR is the False Detection Rate, and the recall (commonly also referred to as completeness) given by:
\begin{eqnarray}
{\rm Recall} = \frac{{\rm TP}}{{\rm TP}+{\rm FN}}. \label{recall}
\end{eqnarray}
The precision is quantifying the reliability of the source extractions, i.e. the likelihood of spurious sources, whereas the recall is giving the probability that a transient will be detected. Calculating the correct number of FN detections is complex as the flux densities of sources will be fluctuating within their uncertainties and hence the source may randomly fall below the detection threshold in the 28 second images. Therefore, we counted the number of FNs assuming no flux density uncertainties and the number of FNs excluding any source in the median image that would be undetected if its flux density had dropped to the 3$\sigma$ lower limit.

In Figure \ref{sourceFinderTest}, the precision and recall are plotted as a function of the flux density of the sources. We find that the precision is consistently $\sim$100\%, therefore we are not dominated by a large number of spurious transient sources. The recall is consistent with $>$90\% for sources above a flux density of $\sim$0.5 Jy, and is in excess of 70\% for flux densities above 0.21 Jy (the sensitivity used in this analysis).  This is a pessimistic lower limit on the recall for new point sources as the source may fall below the detection threshold as the noise level in the image can fluctuate for reasons such as residual calibration errors.

\subsection{Simulations: Transient recovery}

\begin{figure}
\centering
\includegraphics[width=0.48\textwidth]{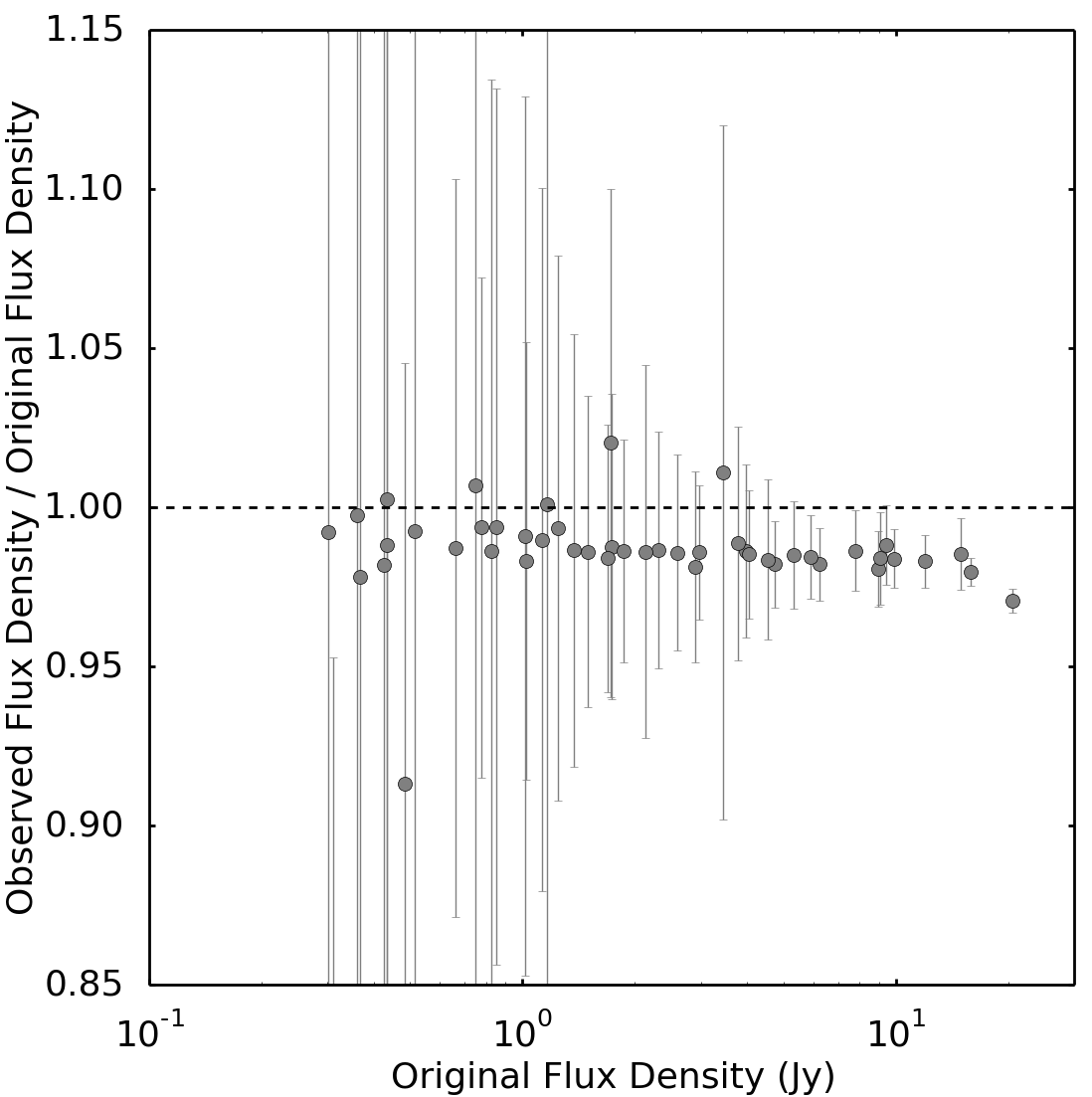}
\caption{The ratio between the observed flux density of the simulated transient to the original flux density when the source was included in the sky model as a function of the original flux density. The black dashed line marks where the two flux densities are equal giving a ratio of 1.}
\label{SimFlxTran}
\end{figure}

Conducting a phase and amplitude self calibration of the dataset using a model image of the field, as described in Section 2.2, leads to significantly improved image quality; the image noise is lower and artefacts surrounding bright sources are reduced. Additionally, this method ensures that the flux density scale is more consistent from image to image across the entire dataset. However, the method has the implicit assumption that the resultant image will be identical to the input image (i.e. the total flux density is constant and is distributed in the same places). If a transient occurs during the observation, or a known source is significantly variable, the output dataset is contrary to this assumption. The expectation is that the calibration step will find the solution which best fits the model and, as long as the majority of the flux density in the model is well known, the transient and variable behaviour will be recovered. The anticipated limit to this method is when the transient flux density significantly exceeds the flux density of the brightest source in the field and then calibration will fail.

In order to confirm that transients within the field are recovered, we simulate a range of transient flux densities within the image. As a transient is a source which is not in the model image used to calibrate the dataset, we can simply simulate a transient by removing a source from the model image (by setting the relevant pixels equal to zero). Using the strategy outlined in Section 2.2, we then apply a phase and amplitude self calibration, using this edited model image, on a previously un-calibrated dataset (observation ID 1061661200). We create 50 new models, by removing 50 sources with a range of flux densities pseudo-logarithmically distributed in the range $\sim$0.3--20 Jy (corresponding to roughly the faintest and the brightest sources in the field) and at random positions within the image, as shown in Figure \ref{SimTran}. Following imaging of the calibrated datasets, we measure the recovered flux density of the simulated transient using the source finding settings used in the transient search (as described in Section 2.6 and Table \ref{table:SFsettings}).

We confirm that all the simulated transient sources are successfully recovered via this calibration and the source extraction method. In Figure \ref{SimFlxTran}, we plot the ratio between the observed and original source flux density as a function of the flux density of the source. The observed flux density is typically within the uncertainties of the flux density measurements, however it is typically lower than the original flux density by a few percent and is proportional to the flux density of the transient source. The only exceptions are for sources where the source finder fits the source as an extended source and may be affected by local structured noise in the image. The reduction in flux density is unsurprising as the calibration assumes that the majority of the flux density in the field is contained in known sources. Therefore, as the brightness of the unknown source becomes a significant contribution of the total flux density in the sky, the calibration is underestimating the total flux density in the field and leads to the reduction in the flux density of the transient source. 

This calibration strategy is expected to fail when the flux density of the transient source dominates the total flux density in the sky model, leading to an extremely poor quality image with few recognisable sources. Therefore, images that significantly fail quality control are a possible signature of a bright transient within the field (although these are likely to have a radio frequency interference origin). The transient surface densities quoted in this paper are for transients that are not significantly brighter than the brightest source in the field, i.e. $\gg$20 Jy. In the event of failed calibrations, recalibration via the transfer of calibration solutions from a calibrator can be used to search for bright transients. As the occurrence of this is expected to be extremely rare, the self calibration strategy used in this paper will give the optimal image quality required for fainter transient sources.

\subsection{Flux density stability between images}

\begin{figure*}
\centering
\includegraphics[width=0.48\textwidth]{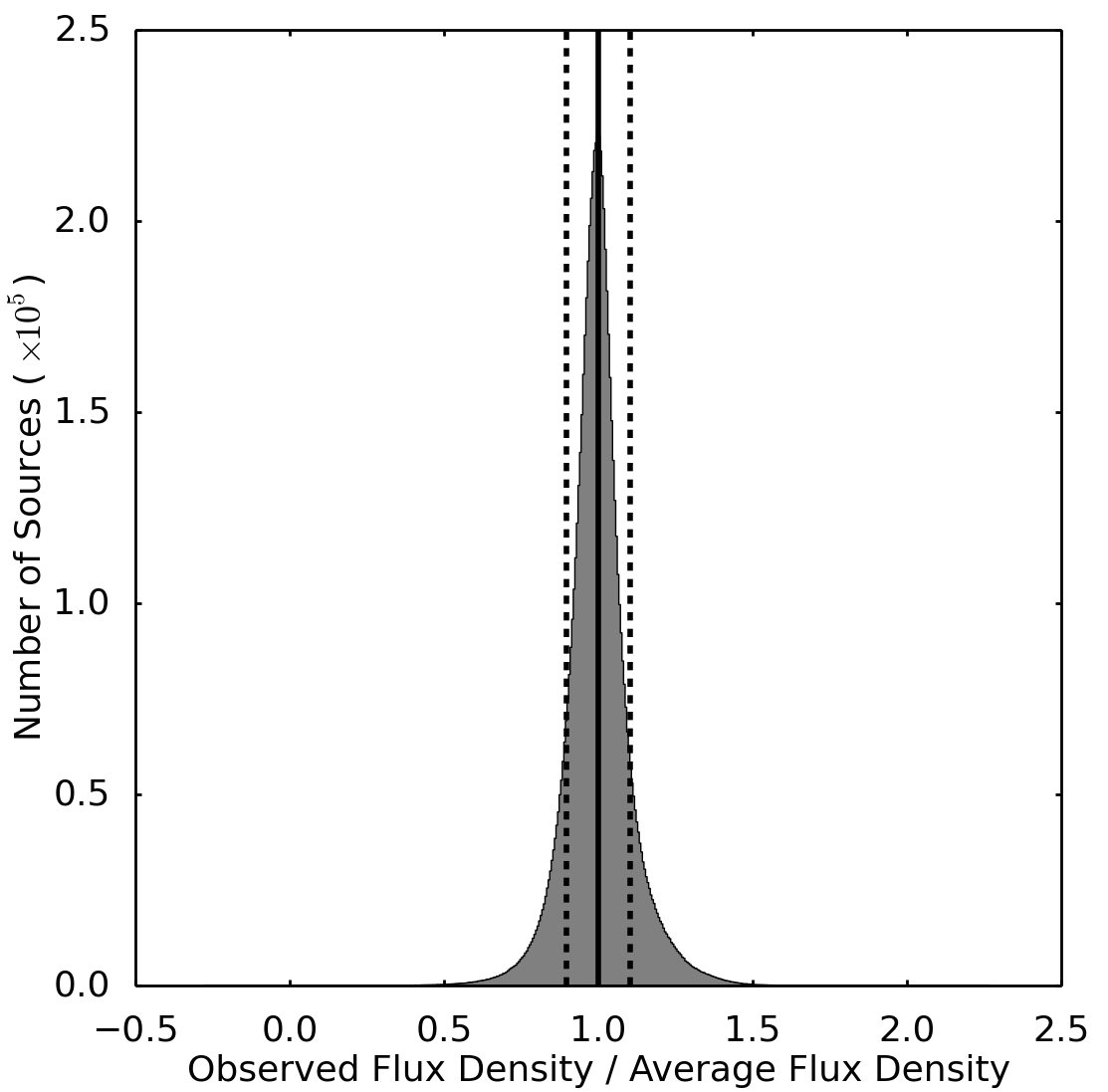}
\includegraphics[width=0.48\textwidth]{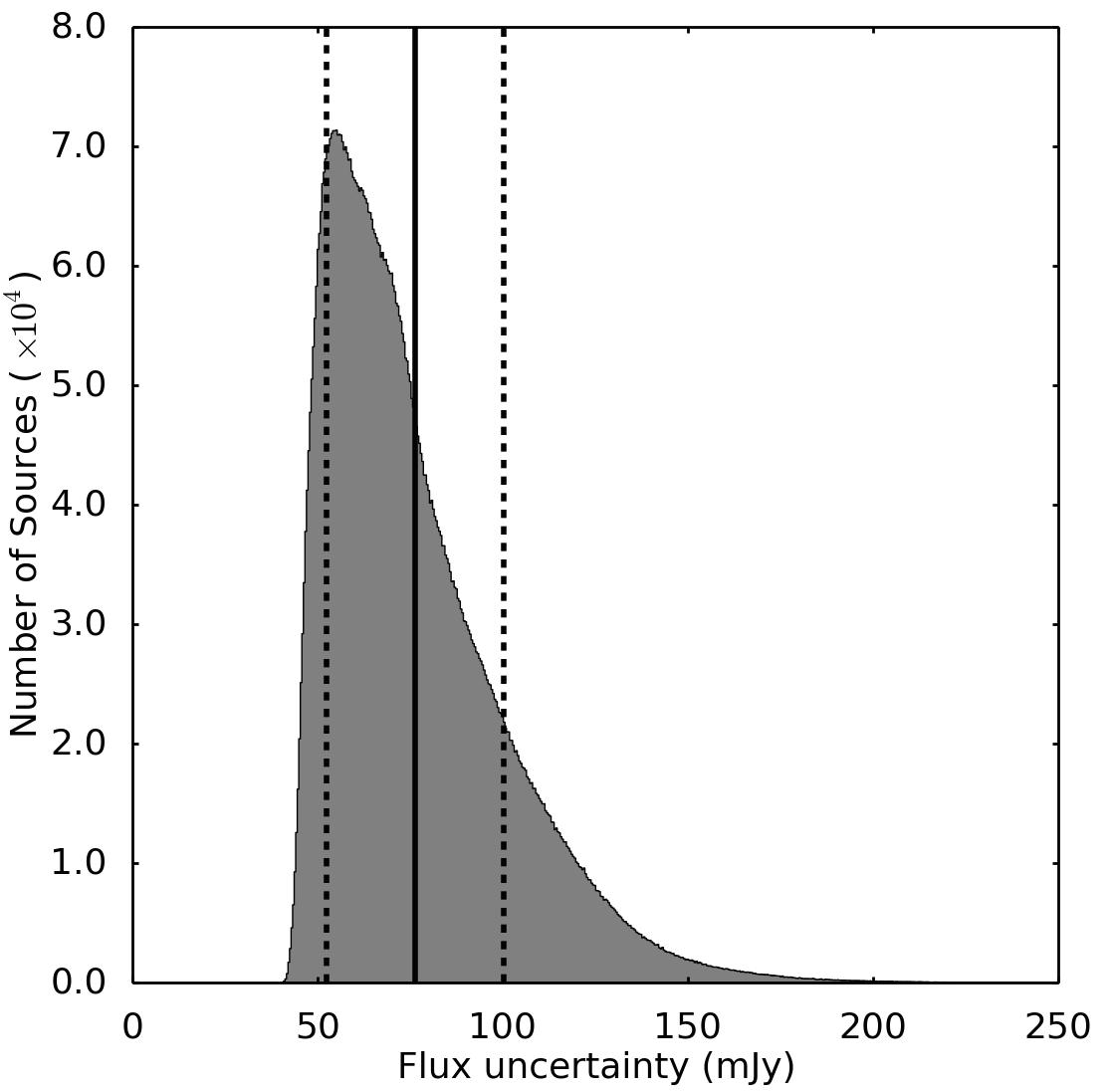}
\caption{Here, we show the histograms for the zenith observations showing the typical distributions of the $\frac{\rm Observed~Flux}{\rm Average Flux}$ (left) and the flux density uncertainties (right). The solid line marks the mean of each distribution with the scattered lines giving the 1$\sigma$ standard deviation.}
\label{FlxQC2}
\end{figure*}

The variability analysis conducted in Section 4 requires a well understood flux density calibration between all the images in the dataset. In this Section, we use all sources in the field with average flux densities in excess of 0.5 Jy, as measured by {\sc TraP} using the monitoring strategy (as described in Sections 2.6 and 4). As stated in Section 2.2, there are residual primary beam effects in the observed flux densities, when moving between different Az-El pointing directions. In this Section and Section 4, we will only consider source variability between detections at the same Az-El pointing directions.

For every source detection, we measure the ratio between the observed flux density and the average flux density of the source and the observed flux density uncertainty. We plot the histograms of these parameters from the zenith Az-El pointing directions in Figure \ref{FlxQC2}. For each distribution we calculate the mean and 1$\sigma$ standard deviation. We find that the extracted source flux densities are typically in excellent agreement with their average flux density with a deviation equivalent to $\sim$10\% of the average flux density at the 1$\sigma$ level. The typical flux density uncertainties are $76\pm24$ mJy, or $\sim2.5\times$ the typical image RMS in the inner part of the image, although this is strongly dependent on radius from the pointing centre due to the decreased sensitivity of the primary beam (the image RMS in the source extraction region varies from a minimum of 23 mJy beam$^{-1}$ to a maximum of 153 mJy beam$^{-1}$) and varies significantly for extended sources.

We find no dependence of the source flux densities on radial position within the image or the image RMS throughout the entire dataset. Therefore, the images from specific Az-El pointing directions have a reasonably consistent flux density scale enabling initial variability searches for the identification of significantly variable sources and is sufficient for this pilot survey.

To confirm the absolute flux density scale, we compare the observed flux densities to the following catalogues:
\begin{itemize}
    \item 180 MHz Murchison Commissioning Survey \citep[MWACS;][]{hurleywalker2014} that fully overlaps with this field. This survey was observed approximately 1 year prior to the start of the observations used in this analysis.
    \item Sydney University Molonglo Sky Survey \citep[SUMSS;][]{mauch2003} at 843 MHz, covering declinations $<-30$ degrees.
    \item VLSSr \citep[][]{cohen2007,lane2014} at 74 MHz, covering declinations $\gtrsim-25$ degrees with sparse coverage of the field in the range $-35$ to $-25$ degrees.
    \item The NRAO VLA Sky Survey \citep[NVSS;][]{condon1998} at 1.4 GHz, covering declinations $> -40$ degrees.\end{itemize}

We cross matched all the sources from each of the Az-El pointing directions separately and compare the ratio between the average source flux density and the catalogue flux density (extrapolated to 182 MHz assuming a typical spectral slope of -0.7). We find the observed flux densities are consistent with the extrapolated catalogue flux densities with an average ratio of 1.1$\pm$0.2 (MWACS), 1.5$\pm$0.5 (SUMSS), 1.4$\pm$0.4 (VLSSr) and 1.6$\pm$0.8 (NVSS). These are all consistent within the 1$\sigma$ uncertainties. However we note that these ratios are all larger than unity, which is most likely a systematic offset caused by uncertainties in the primary beam model and the difference in elevation between the observation of 3C444 (used to calibrate the flux scale) and the observations.

\section{Transient Analysis}

This dataset constitutes one of the largest sky areas, with a field of view of $\Omega = 452$ square degrees, surveyed with excellent sensitivities on a very wide range of timescales (30 seconds -- $\sim$1 year) at low radio frequencies. The field is well off the Galactic plane and hence likely to be dominated by extragalactic sources. This enables us to place tight constraints on the surface density of extragalactic low-frequency radio transients.

We processed all the images with {\sc TraP} utilising the transient search source finder parameters given in Table \ref{table:SFsettings}. To aid with source association, we use a systematic position uncertainty of 270 arcsec, corresponding to 5 pixels, which is added in quadrature to the position uncertainties for each source. As we are only searching for transient sources in this section, we can assume that they will not be detected in the deep median image (see Section 2.5 for more details) so we forced {\sc TraP} to process the median image first to create a deep source catalogue for source association. Therefore, any sources identified as new sources from the subsequent, shallower 30 second images are all transient candidates. As we focused on the detection of new sources and not their variability parameters in this section, we were able to subdivide the dataset by time into smaller, more manageable, chunks for processing. 

The source finder settings used will occasionally lead to large elongated Gaussians fitted to noise structures in the image. These fitting errors can be easily mitigated against as we expect all point transient sources to be roughly circular in these images (the restoring beam is close to circular), so we rejected all fitted sources with Gaussian shapes with $\frac{\rm major~axis}{\rm minor~axis}>2$.

Although the self-calibration strategy has reduced the number of artefacts around bright sources, there remained an over density of new source detections around the bright sources. Visual inspection of a randomly chosen sample confirmed these are caused by side-lobes of the bright sources. Due to this, we rejected all candidates that occurred within 0.4 degrees of a source with flux densities in excess of 8 Jy, leading to a reduction in surveyed area of $\sim$3 square degrees.

A further over-density of transient candidates occurred on the source extraction boundary at a radius of 12 degrees. This is caused by the source finder not modelling the RMS noise in the region beyond the source extraction region. We visually inspected a randomly chosen sample of these candidates and, by increasing the source finder radius to better model the RMS noise in this region, showed that these were not significant sources. Therefore, we rejected all candidates occurring within 0.2 degrees of the edge of the source extraction region, leading to a reduction in surveyed area of $\sim$15 square degrees.

Following this, we overlaid the transient candidates on the median image and rejected all candidates that had a counterpart in the median image\footnote{We note that these may be variable sources and will investigate further in future analysis.}. We identified 3 candidates requiring further analysis and visually inspected the detection image. One of the candidates was rejected following visual inspection, as it was consistent with an artefact that was a deconvolution error of a source in close proximity to a $\sim$15 Jy source.

We developed a number of further tests using one of the most convincing of the 2 remaining candidates (with duration $\sim$30 seconds, detected at 6.45$\sigma$):
\begin{enumerate}
    \item The images were processed using different source finder settings and a different source finder, {\sc Aegean} \citep{hancock2012}, to confirm the detection significance.
    \item We re-imaged the region with a range of different imaging parameters; such as changing pixel scale, weighting, image phase centre, and UV range.
    \item We produced new images on a range of additional timescales (2 minutes, 10 seconds and 4 seconds).
    \item The transient candidate remained detected following these tests, so we also processed the observation using the de-dispersion pipeline in Miriad developed by \cite{tingay2015}, resulting in a faint detection of the source in the Miriad images but a non-detection of dispersed signals.
    \item We conducted an image subtraction using an image with the identical local sidereal time (LST) from the previous night. The transient candidate was not significantly detected in the subtracted image, suggesting the source is related to a correlated noise artefact (see Section 2.4).
    \item A median image was created using all the images in the dataset with the identical LST and confirmed the presence of a noise peak at this position for this LST. The transient candidate had a flux density in excess of the noise feature in the median image however, given the noise feature and the low SNR, this is an unconvincing source.
\end{enumerate}
The remaining transient candidate was also consistent with a noise peak at a specific LST and, hence, is unconvincing.

The processing strategy outlined in this Section may fail to detect bright long duration ($\sim$months) transient sources on year timescales, as they could leave a residual source in the median image that could be detected by the source finder (for instance a 300 mJy transient source with a duration of 2 months could potentially lead to a $\sim$50 mJy source, i.e. a $\sim$6$\sigma$ detection, in the median image). However, the targeted population of transient sources will not have a counterpart in the existing radio catalogues covering this region. Therefore, we cross-match all the sources detected in the median image with the NVSS catalogue (the only catalogue covering the entire region). The peak flux densities for sources that were not detected in NVSS were extrapolated to 1.4 GHz to determine if we expected to detect them. Eighteen faint, uncatalogued radio sources were identified in the median image, however they are not expected to be detectable in previous surveys of this region. We used the monitoring capability in {\sc TraP} (see Sections 2.6 and 4 for more details) to obtain a light curve at the position of these faint sources to confirm that they were not residual sources from bright transients. Using this strategy, we identified no long duration bright transient sources and analysis is ongoing to determine if these sources are variable.

\subsection{FRB limits}

\begin{figure}
\centering
\includegraphics[width=0.47\textwidth]{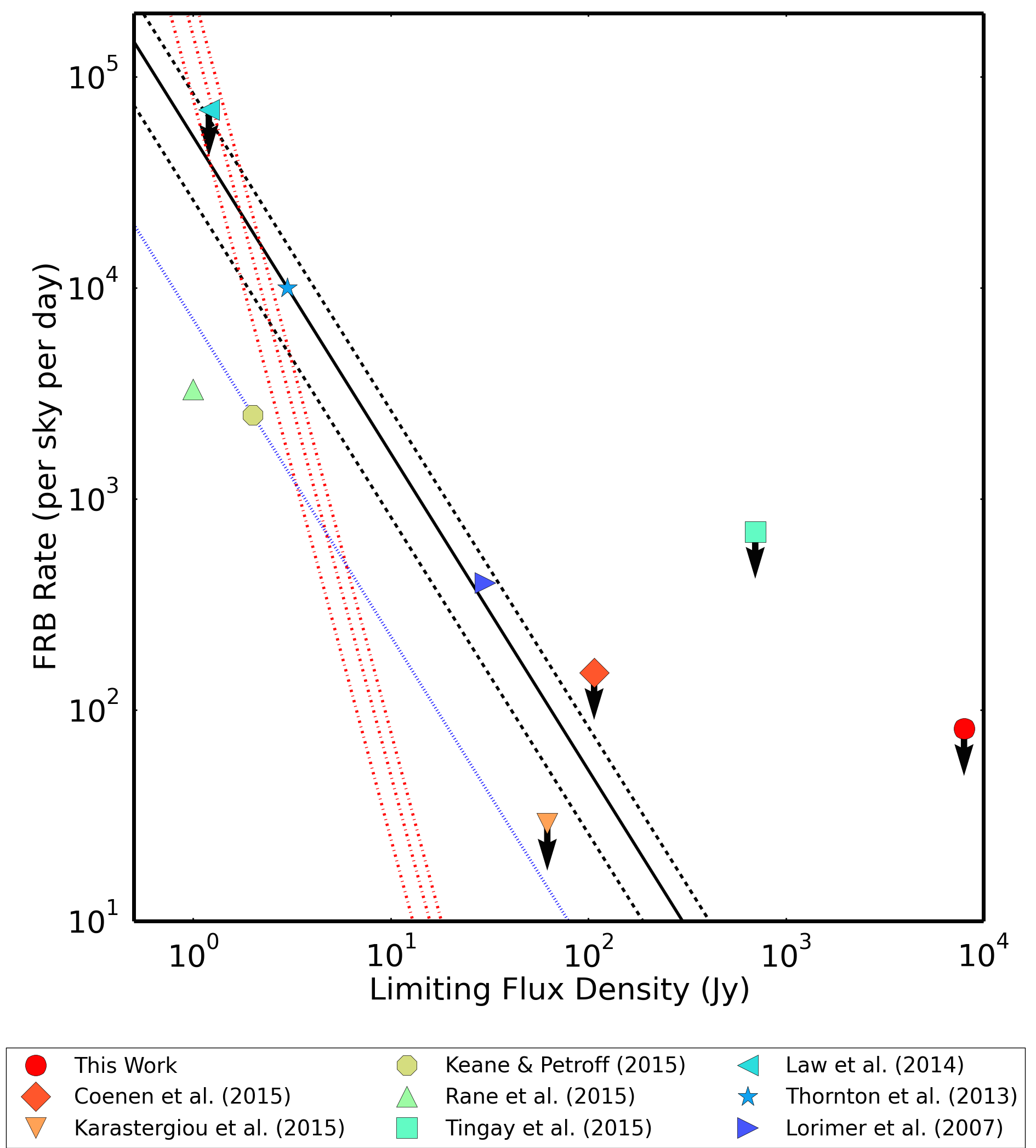}
\caption{This plot, based on plot from \citet{coenen2014}, shows the FRB rate per sky per day as determined from the surveys given in Table \ref{table:FRBsurveys}, assuming a flat spectral slope. This plot assumes that the only limiting factor is the survey sensitivity. Symbols are as in the legend and arrows denote upper limits. Assuming a cosmological population, where the only constraining factor is the sensitivity of the survey (although note this is not a valid assumption for FRBs, see the text for further details), the data points are expected to be consistent with a straight line of slope -1.5. We plot a solid black line representing this and normalised to \citet{thornton2013} with the uncertainties bounded by the dashed black lines. For reference, we also plot this for the lower rate given by \citet{keane2015} using a blue dotted line. Finally the red dash-dotted lines represent the significantly steeper and lower rate model proposed by \citet{macquart2015}.}
\label{FRBrates}
\end{figure}

\begin{table*}
\centering
\caption{The number of FRBs expected to be detected in the 28 second images as a function of the spectral index of the FRB, assuming the observed rates from \citet{thornton2013} and 75\% lower rates consistent with the analysis by \citet{keane2015}. } \label{table:predictFRB} 
\begin{tabular}{|c|c|c|c|c|}
\hline
Spectral index & \multicolumn{2}{|c|}{\citet{thornton2013}}     & \multicolumn{2}{|c|}{\citet{keane2015}} \\
               & Number predicted & Null detection probability & Number predicted & Null detection probability \\
\hline
-2             & 27.1             & $1.8\times10^{-12}$        & 6.8              & $1.2\times10^{-3}$         \\
-1             & 4.8              & $8.2\times10^{-3}$         & 1.2              & $3.0\times10^{-1}$         \\
0              & 0.7              & $4.8\times10^{-1}$         & 0.2              & $8.3\times10^{-1}$         \\

\hline
\end{tabular}
\end{table*}

\begin{table*}
\centering
\caption{The current rate constraints for FRBs. $^*$ The peak flux density sensitivity given by \citet{rane2015} is converted to an observed flux density for a pulse that is on for $\sim$10 ms \citep[the average duration observed by][]{keane2015}.} \label{table:FRBsurveys} 
\begin{tabular}{|c|c|c|c|c|c|c|}
\hline
Survey & Frequency & Sensitivity & Rate & DM Range & Method & Citation \\
  & (MHz) & (Jy) & (/day/sky) & (pc cm$^{-3}$) & & \\
\hline
MWA               & 182 & 7980 & $<82$ & $<$700 & 28 second images & This Work \\
\hline
MWA               & 156  & 700 & $<700$           & 170--675  & 2 second de-dispersed images & \citet{tingay2015} \\
ARTEMIS           & 145  & 62   & $<29$            & $<$320    & High time resolution         & \citet{karastergiou2015} \\
Parkes (Bayesian, All) & 1400 & 1$^*$  & $3.3^{+5}_{-2.5} \times10^{3}$       & $\sim$2000 & High time resolution         & \citet{rane2015} \\
Parkes (9 FRBs)   & 1400 & 2    & $\sim2500$       & 375--1103 & High time resolution         & \citet{keane2015} \\
LOFAR             & 142  & 107  & $<150$           & 2--3000   & High time resolution         & \citet{coenen2014} \\
VLA               & 1400 & 1.2  & $<7\times10^{4}$ & 0--3000   & High time resolution         & \citet{law2014} \\
Parkes (4 FRBs)   & 1400 & 3    & $1^{+0.6}_{-0.5} \times10^{4}$ & 553--1103 & High time resolution         & \citet{thornton2013} \\
Parkes (1 FRB)    & 1400 & 30   & $\sim400$        & 375       & High time resolution         & \citet{lorimer2007} \\
\hline
\end{tabular}
\end{table*}

This dataset, with an snapshot timescale of 28 seconds, has a very good cadence for the detection of FRBs in the image plane (without de-dispersion) at 182 MHz. By scaling from known rates for FRBs detected at 1.4 GHz, we can predict the number of FRBs we expect to observe in the 10,122 images included in this analysis.

Using the methods presented in \cite{trott2013}, we predicted the number of FRBs that we expect to detect in the 28 second images, with no image plane de-dispersion, as a function of a range of spectral slopes assuming that FRBs are a standard candle. In a 28 second image we are able to probe dispersion measure (DM) values up to 700 pc cm$^{-3}$, within this component we estimate a Galactic DM of $\sim$10 pc cm$^{-3}$ (field is well off the Galactic plane, so the Galactic component is low) and a host galaxy component of 100 pc cm$^{-3}$. The rates predicted are normalised to the whole sky rate of FRBs observed by \cite{thornton2013}. For the \cite{thornton2013} rates, we rule out spectral indices $\le$-1 at $>$95\% confidence, while for the 75\% lower rates reported by \cite{keane2015} we are able to rule out spectral indices $\le$-2 at $>$95\% confidence.

In this dataset, we expect all FRBs to only be detected in 1 image, as their flux densities will likely be too low to be detected if spread over multiple images. To determine if there are any FRB candidates in this dataset, we queried all of the new sources detected by {\sc TraP} to identify all sources which are only detected in 1 image and that were not found in the first, deep median image. This resulted in a small number of candidates that were visually inspected using both the median image, the detection image and other 30 second images close by in time. The majority of the remaining sources had visible counterparts in the median image, some of which are related to failed source associations and others are candidate variable sources warranting future investigation but do not meet our requirements for candidate FRBs.

Our non-detection of any FRB candidates can place tight constraints on the rates of FRBs at low radio frequencies. Following the method adopted by \cite{trott2013}, where we are in the regime that the snapshot duration is greater than the signal width, we can estimate the minimum FRB flux densities $S_{\rm FRB,min}$ that we are sensitive to using:
\begin{eqnarray}
S_{\rm FRB,min} = S_{\rm min,28s} \left( \frac{\Delta t}{w} \right), \label{eqn:limFlx}
\end{eqnarray}
where $S_{\rm min,28s} =0.285$ Jy is the sensitivity in 1 snapshot image, $\Delta t = 28$ s is the snapshot integration time and $w$ is the intrinsic width of the FRB. For consistency with \cite{tingay2015}, we assume that the intrinsic width is 1 ms. Therefore, this experiment is sensitive to FRBs with flux densities in excess of 7980 Jy. We can estimate an upper limit on the rate of FRBs per sky per day observable using the standard transient rate for the 28 second snapshot rate, $3\times10^{-6}$ deg$^{-2}$ (as calculated in Section 3.2). To observe the equivalent of 1 day would require 3085 snapshots of 28 second integration time, therefore our whole sky FRB rate ($\rho_{\rm FRB}$) is:
\begin{eqnarray}
\rho_{\rm FRB} (S_{\rm FRB} > 7980~Jy) < 82 /{\rm sky}/{\rm day} \label{eqn:FRBrate}
\end{eqnarray}
In Table \ref{table:FRBsurveys} and Figure \ref{FRBrates} \citep[adapted from ][ assuming a flat spectral slope]{coenen2014}, we show this FRB rate in comparison to previous surveys at a range of frequencies, assuming FRBs have a flat spectrum. Here, we assume that FRBs are a standard candle and the observed FRB population have been shown to be consistent with this \citep{dolag2015}. Assuming a flat spectrum and a cosmological population and using
\begin{eqnarray}
N \propto S^{-\frac{3}{2}}, \label{eqn:cosmologicalPop}
\end{eqnarray}
where N is the number of transients and S is the flux density of the transient, we can determine that the rates we obtain are broadly consistent with the rate obtained by \cite{tingay2015}. However, we note that our rate limit is higher than that expected when extrapolating the \cite{thornton2013} rate to our sensitivity and is unconstraining for flat spectral slopes. Additionally, a recent calculation by \cite{keane2015} has shown that the observed FRB rate may be 4$\times$ lower than that determined by \cite{thornton2013}. Recently \cite{macquart2015} have postulated that the rate of FRBs does not follow that of a standard cosmological population, explaining the lack of FRBs at low Galactic latitudes, and instead is given by:
\begin{eqnarray}
N \propto S^{-\frac{7}{2}} \label{eqn:cosmologicalPop2}
\end{eqnarray}
A $S^{-\frac{7}{2}}$ distribution is no longer considering a cosmological population or a standard candle, i.e. this would imply that either the population or the luminosity is strongly dependent upon the redshift. Alternatively, there may be further selection effects that have not been accounted for. Additionally, they show that the whole sky rates will be a factor of 3 lower than the observed rates, therefore normalising by a third of the \cite{thornton2013} rate, we plot this constraint in Figure \ref{FRBrates} and our upper limits are consistent with this underlying population. We note that the rates determined by \cite{lorimer2007} are significantly higher than this model, even when the expected uncertainties on this rate are taken into account.

Figure \ref{FRBrates} assumes that only the survey sensitivity is required to account for the volume probed by each of the surveys. However, for FRBs this is not strictly the case as it also strongly depends on the DM range searched over. The DM search range constrains the volume that can be searched for FRBs, irrespective of their luminosities. In this analysis, the 28 second images produced can be used to probe DMs up to 700 pc cm$^{-3}$, whereas other surveys can exceed DMs of 1000 pc cm$^{-3}$ or can be much lower than this. The DM can be converted to a redshift (z) using the relationship \citep[e.g.][]{ioka2003,inoue2004,lorimer2007,karastergiou2015}\footnote{See also \cite{dolag2015} for an alternative method to constrain the redshift.}:
\begin{eqnarray}
DM \approx 1200~z~{\rm cm}^{-3}~{\rm pc}. \label{eqn:DMzRelation}
\end{eqnarray}

However, before conversion to a redshift, we want to remove the Galactic component of the DM and also a contribution from the host galaxy of the FRB. The host galaxy contribution is unknown and, for consistency with \cite{karastergiou2015}, we assume it is 100 pc cm$^{-3}$. Using the model produced by \cite{cordes2002}\footnote{Using the web interface here: http://www.nrl.navy.mil/rsd/RORF/ne2001/} we find a DM of 29 pc cm$^{-3}$, however we note that this model is based on the extrapolation of DM measurements from pulsars and only one of the pulsars used is within our field\footnote{\cite{gaensler2008} finds a DM of 25 pc cm$^{-3}$ for similar Galactic latitudes but in different directions to this field.}. This is a small amount relative to the DM searched, unsurprising due to the Galactic latitude of the target field (centred on the Galactic co-ordinates l,b: 30.6,$-$78.5 degrees),  and is therefore negligible in our analysis. Therefore, the DM component that is expected to be from the intergalactic medium (IGM) is $\sim$600 pc cm$^{-3}$, corresponding to a redshift of $\sim$0.5. 

In conclusion, we find a rate of $<$82 FRBs per sky per day that are brighter than 7980 Jy at a frequency of 182 MHz out to a maximum redshift of $\sim$0.5. We show our non-detection is consistent with the lower rates calculated by \cite{keane2015}, if FRBs have a spectral slope $>-1$, or the non-standard cosmological population suggested by \cite{macquart2015}.

\subsection{Transient Rates}

\begin{table*}
\centering
\caption{The low frequency radio surveys undertaken to date ($<$1 GHz) and the current surface density constraints.} \label{table:surveys} 
\begin{tabular}{|c|c|c|c|r|}
\hline
Survey & Frequency & Sensitivity & Timescale & Surface density (minimum timescale) \\
  & (MHz) & (Jy) & & (deg$^{-2}$)  \\
\hline
This Work             & 182    & 0.285  & 28 seconds &  $<6.4\times10^{-7}$ \\
This Work             & 182    & 0.285  & 5 minutes  &  $<6.6\times10^{-6}$ \\
This Work             & 182    & 0.285  & 10 minutes &  $<1.1\times10^{-5}$ \\
This Work             & 182    & 0.285  & 1 hour     &  $<2.5\times10^{-5}$ \\
This Work             & 182    & 0.285  & 2 hours    &  $<9.5\times10^{-5}$ \\
This Work             & 182    & 0.285  & 1 day      &  $<1.2\times10^{-4}$ \\
This Work             & 182    & 0.285  & 3 days     &  $<2.4\times10^{-4}$ \\
This Work             & 182    & 0.285  & 10 days    &  $<3.9\times10^{-4}$ \\
This Work             & 182    & 0.285  & 30 days    &  $<9.5\times10^{-4}$ \\
This Work             & 182    & 0.285  & 90 days    &  $<3.3\times10^{-3}$ \\
This Work             & 182    & 0.285  & 1 year     &  $<6.6\times10^{-3}$ \\
\hline
\citet{stewart2015}     & 60     & 36.1   & 30 seconds -- 4 months &  $<4.1\times10^{-7}$ \\
\citet{stewart2015}     & 60     & 7.9    & 4 minutes -- 4 months  &  $1.4\times10^{-5}$ \\
\citet{obenberger2015} & 38     & 1440   & 5 seconds         & $<1.9\times10^{-11}$ \\
\citet{bell2014}       & 154    & 5.5    & minutes -- 1 year & $<7.5\times10^{-5}$  \\
\citet{carbone2014}    & 150    & 0.5    & minutes -- months & $<1\times10^{-4}$    \\
\citet{cendes2014}     & 149    & 0.5    & minutes -- months & $<2.2\times10^{-2}$  \\
\citet{jaeger2012}     & 325    & 0.0021 & 1 day -- 3 months & $1.2\times10^{-1}$   \\
\citet{bannister2011}  & 843    & 0.014  & days -- years     & $1.3\times10^{-2}$   \\
\citet{lazio2010}      & 74     & 2500   & 300 seconds       & $<9.5\times10^{-8}$  \\
\citet{hyman2009}      & 235,330 & 0.03  & days -- months  & $3.4\times10^{-2}$  \\
\hline
\end{tabular}
\end{table*}

\begin{figure*}
\centering
\includegraphics[width=\textwidth]{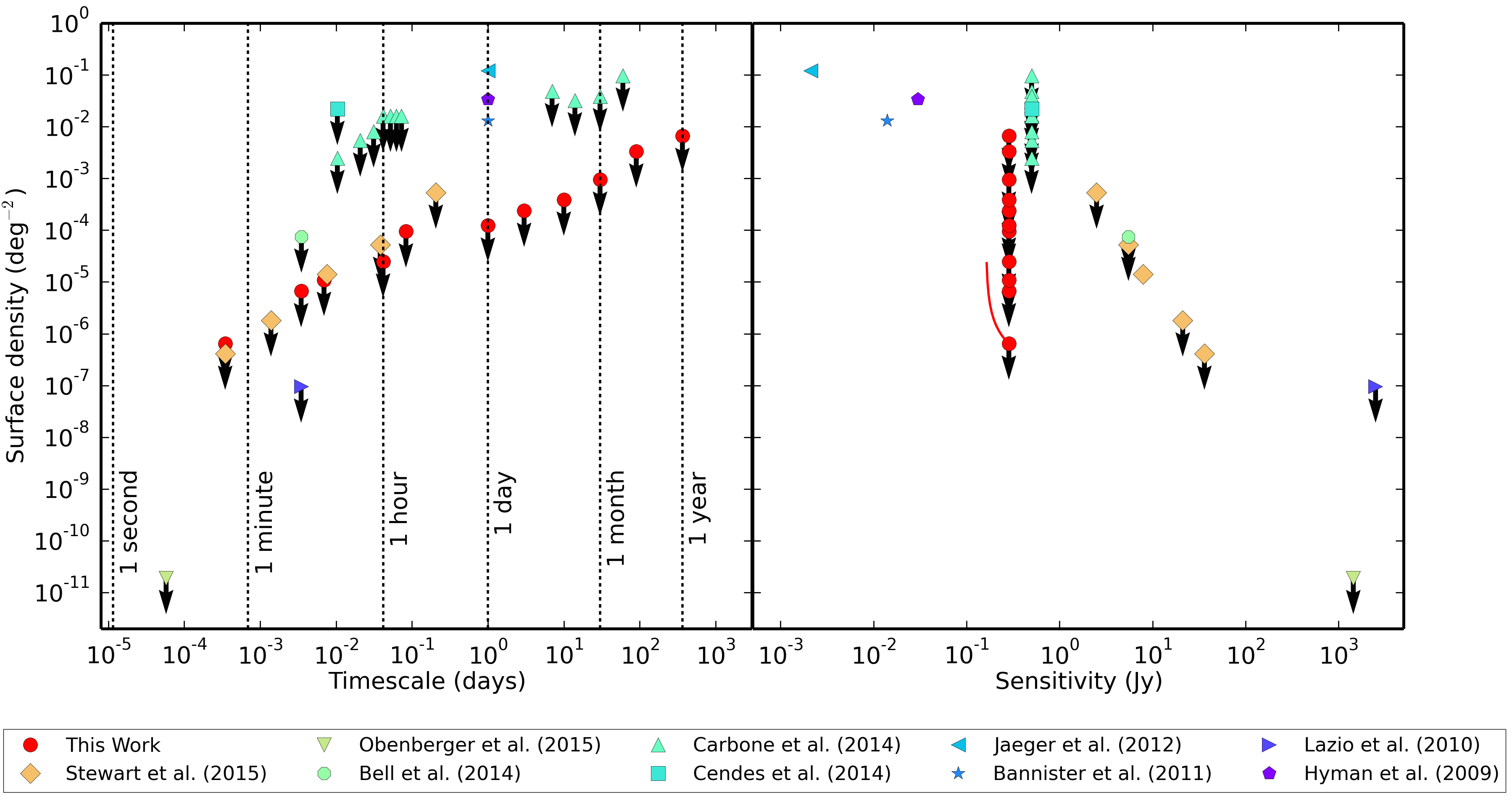}
\caption{The observed transient surface densities plotted as a function of the timescales probed (left) and the limiting flux density (right). The different surveys are given in the legend and down arrows denote upper limits. The red solid line represents the typical rate constraints taking into account the differing sensitivity across the images due to the primary beam response (assuming a Gaussian primary beam), starting at a radius of 2 degrees and ending at the source extraction radius of 12 degrees.}
\label{rates}
\end{figure*}

In addition to the non-detection of FRB candidates (as described in Section 3.1), no convincing transient sources were detected in this dataset. We can calculate the standard transient surface density limit using Poisson statistics via:
\begin{eqnarray}
P = \exp^{-\rho (N-1)\Omega}, \label{eqn:rate}
\end{eqnarray}
where $(N-1)\Omega$ is the total area surveyed by $N$ snapshots of a field each with an area of $\Omega$, $\rho$ is the surface density limit and P is the confidence interval. Following \cite{bell2014}, we utilise $P=0.05$ to give a 95\% confidence limit.

The sensitivity to transients depends upon the location within the image, such as the increase in RMS noise with radius due to the decreasing sensitivity of the primary beam \citep[e.g.][]{croft2013,bell2014}, and on the specific images used as some are of higher quality \citep[e.g.][]{carbone2014}. Here, we can characterise the area surveyed for a given sensitivity by assuming that the sensitivity as function of the position (radius, $r$) within the primary beam, Sensitivity$(r)$, can be approximated as a 1-dimensional Gaussian distribution given by:
\begin{eqnarray}
{\rm Sensitivity}(r) = 6 ~{\rm RMS_{\rm c}} \exp \left(\frac{r^2}{2 ~{\rm HWHM}^2} \right) ~{\rm mJy}, \label{eqn:primaryBeamApprox}
\end{eqnarray}
where RMS$_{\rm c}$ is the RMS in the central 100 pixels of the image and the factor of 6 is because we extract sources which are $6\sigma$ above the detection threshold. By measuring the RMS in the central 100 pixels of all the images, we find an average RMS$_{\rm c} = 27$ mJy beam$^{-1}$. For the full source extraction region, with a radius of 12 degrees and an area of 452 square degrees, we have a sensitivity of $\sim$285 mJy. We note this is still an approximation to the sensitivity as the primary beam response may not take a simple Gaussian shape and does not account for other causes of variation in RMS throughout the field. However, due to the large number of images utilised in this analysis, a detailed characterisation of the sensitivities and surface densities \citep[such as that conducted by][]{bell2014}, for all the surface densities on each of the different timescales probed, would take a disproportionate amount of compute time and not significantly affect the results presented.

In this dataset, we extract sources within a radius of 12 degrees which are $6\sigma$ above the detection threshold corresponding to a detection limit of $\le$0.285 Jy (see Equation \ref{eqn:primaryBeamApprox}), with a field of view of $\Omega = 452$ square degrees. On the shortest timescale, 28 seconds, we used $N=10,122$ images which corresponds to a surface density of $<6.4\times10^{-7}$ deg$^{-2}$. We compare our observations to the other transient surveys conducted at low radio frequencies to date in Table \ref{table:surveys}.

Using the method presented in \cite{carbone2014}, where only statistically independent images are used to calculate the rates for a specific transient timescale, the transient surface density limit is calculated for a range of unique timescales probed by this dataset. In Figure \ref{rates}, we plot the surface densities obtained for each of the timescales in comparison to the existing surveys at low frequencies. This highlights the orders of magnitude decrease in the transients surface density limits on a range of timescales that this survey provides. The only surveys with surface densities on faster timescales or lower surface density constraints are for flux density sensitivities that are orders of magnitude higher than this survey. We note that we do not have sensitivity on timescales between $\sim$2 hours and $\sim$1 day due to the EoR observing strategy. Also, as the observations occur when the field is optimally located in the observable sky, we have no sensitivity for transients between timescales of $\sim$3 months to $\sim$1 year. In addition to these gaps in sensitivity, we note that we will also have some sensitivity to transients with timescales $<$28 seconds where very short lived coherent transients from Galactic sources, such as pulsars, may be anticipated (the sensitivity to these transients is dependent upon the flux density and duration of the transient, see Equation \ref{eqn:limFlx}). Finally, we have not probed timescales $\gtrsim$1 year where we expect to observe the longer duration synchrotron sources at this observing frequency. These parameter spaces remain to be explored in the future. 

In Figure \ref{rates} we plot the standard surface densities versus the sensitivity (flux density detection limit) for the sample of low frequency radio transient surveys, the published surface densities from higher radio frequencies are plotted for reference. This survey is typically greater than an order of magnitude more sensitive than previous studied or, conversely, orders of magnitude more constraining for a given sensitivity. To further constrain these rates, future surveys will require significantly increased sensitivity (for instance via SKA-low) or have at least an order of magnitude increase in surveyed area (e.g. via commensal observations).

To date, few of the transients surveys $<$1 GHz have detected transient sources, with most detections at an order of magnitude higher sensitivity \citep[e.g.][]{jaeger2012,bannister2011,hyman2009} on the days--months timescales making them unlikely to be detected in this survey. \cite{stewart2015} have identified a bright transient source at 60 MHz on the minutes timescale using LOFAR \citep[see also ][]{fender2015}. Assuming a flat spectrum, with two orders of magnitude improvement in sensitivity, this survey would be naively predicted to detect hundreds of these transients. As no transients were detected on this timescale, we investigate the implications for the \cite{stewart2015} transient source. Assuming these sources are a standard candle distributed isotropically, we can scale the sensitivity of this survey to that of \cite{stewart2015} via Equation \ref{eqn:cosmologicalPop}. In Table \ref{table:transPred} we predict, for a range of spectral indices, the number of transients expected in this survey and calculate the probability of null detection. Assuming the surface density is correct, we can rule out spectral indices $\ge -2$ at $>$95\% confidence. We conclude that the most likely scenarios are: the transient surface density is much lower than observed by \cite{stewart2015} and/or the spectral index of these transients is very steep ($<-2$). A steep spectral index may be consistent with this emission being from coherently emitting sources such as pulsars. This is consistent with the very steep spectrum $<-4.$ proposed by \cite{stewart2015}.

Finally, we note that this field is close to the Galactic poles and does not constrain the Galactic population of transient sources. This will be best determined by up-coming whole sky transient surveys.

\begin{table}
\centering
\caption{The number of transients expected to be detected on the minutes timescale, as a function of the spectral index of the transient, scaled from the transient detection by \citet{stewart2015}} \label{table:transPred} 
\begin{tabular}{|c|c|c|c|c|}
\hline
Spectral index & Number predicted & Null detection probability \\
\hline
-3             & 1.0         & $3.7\times10^{-1}$ \\
-2             & 5.2         & $5.5\times10^{-3}$ \\
-1             & 27.6        & $1.0\times10^{-12}$ \\
0              & 146         & 0 \\ 
\hline
\end{tabular}
\end{table}

\section{Variability search}

\subsection{Method}

For this section, we put all the images through {\sc TraP} utilising the monitoring strategy described in Section 2.6 and monitor the variability of sources detected in the median image with flux densities in excess of 0.5 Jy. As stated previously, there are residual primary beam issues when comparing images at different Az-El pointing directions so we process each pointing direction separately. Variable candidates are then compared between each of the pointing directions.

{\sc TraP} measures two key variability parameters for every unique source in the dataset. The first parameter is the reduced weighted $\chi^{2}$ at a given observing frequency, $\eta$, given by:
\begin{eqnarray}
\eta = \frac{N}{N-1}\Biggl(\overline{w\,{I}^2}-\frac{\overline{w\,I}^2}{\overline{w}}\Biggr) \label{eqn:eta_nu} 
\end{eqnarray}
where $N$ is the number of datapoints, $I_{i}$ is the flux density of a datapoint, $\overline{w} = \frac{1}{N}\sum_{i=0}^N w_{i} \equiv \sum_{i=0}^N \frac{1}{\sigma_{i}^2}$, $\sigma_{i}$ is the error on the $i$-th flux density measurement and over bars represent the mean values \citep[the full derivation of this from the standard reduced weighted $\chi^{2}$ is given in][]{swinbank2015}. The second parameter is the coefficient of variation (also known as the modulation index) at each observing frequency, $V$, given by:
\begin{eqnarray}
V =\frac{s}{\overline{I}}, \label{eqn:V_nu} 
\end{eqnarray}
where $s$ is the standard deviation of the observed flux densities. These parameters are measured for each time step that the source is observed. In the following analysis, we focus on the variability parameters for each unique source from the final time step.

\begin{figure*}
\centering
\includegraphics[width=0.7\textwidth]{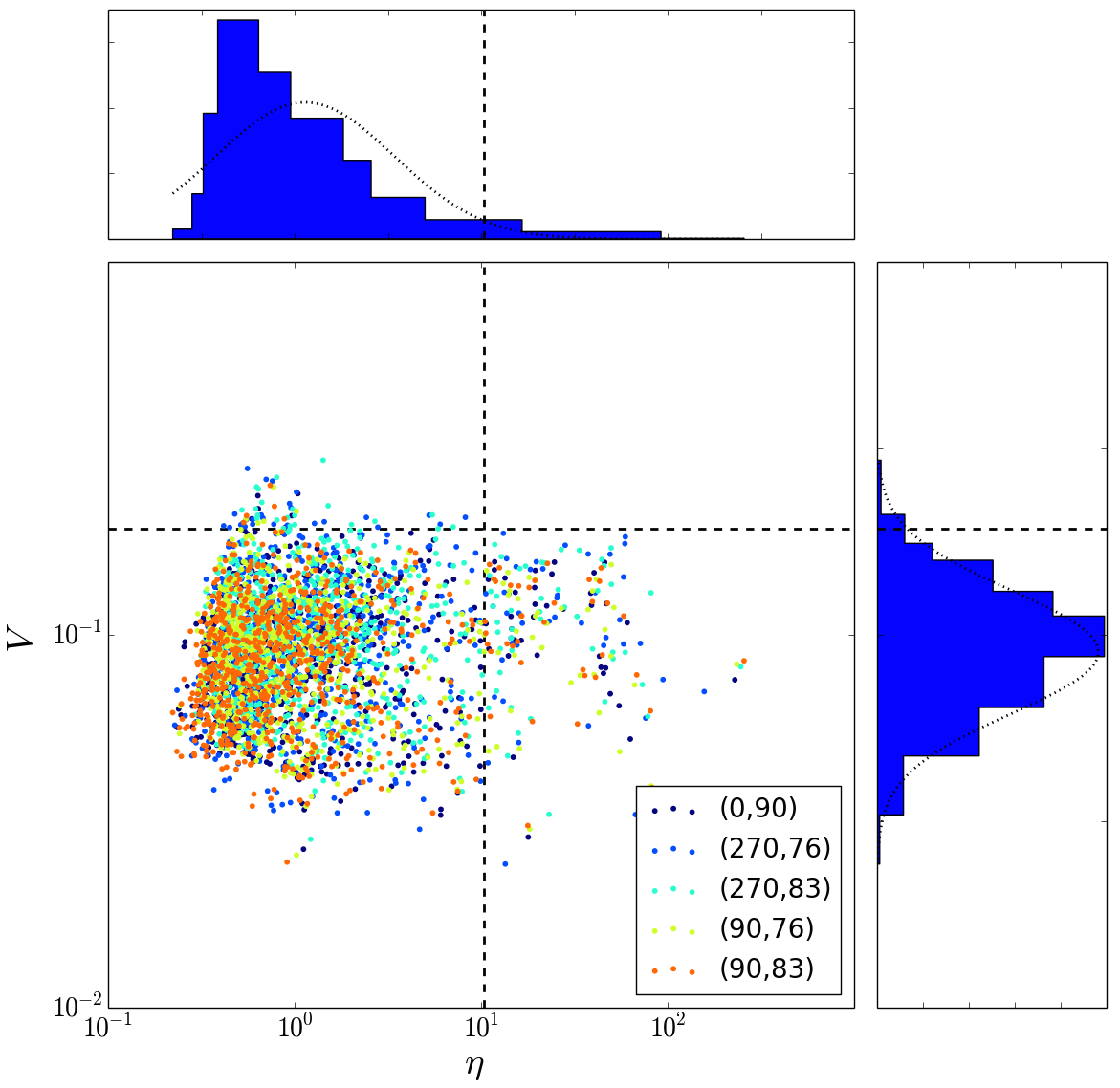}
\caption{This plot shows the two variability parameters: the reduced weighted $\chi^{2}$, $\eta$, and the coefficient of variation, $V$ (given by equations \ref{eqn:eta_nu} and \ref{eqn:V_nu} respectively) for all the sources tracked in the variability search. The colour scheme corresponds to the five separate unique azimuth-elevation phase centres, as given in Table \ref{table:calibrators} and the legend. In the top and right panels, we plot histograms of the two variability parameters and fit them with a Gaussian distribution (dotted line). We utilise a 2$\sigma$ threshold on both parameters, represented by the black dashed line. Significantly variable sources would reside in the top right corner bounded by the two thresholds.}
\label{variPlt}
\end{figure*}

\begin{figure*}
\centering
\includegraphics[width=0.7\textwidth]{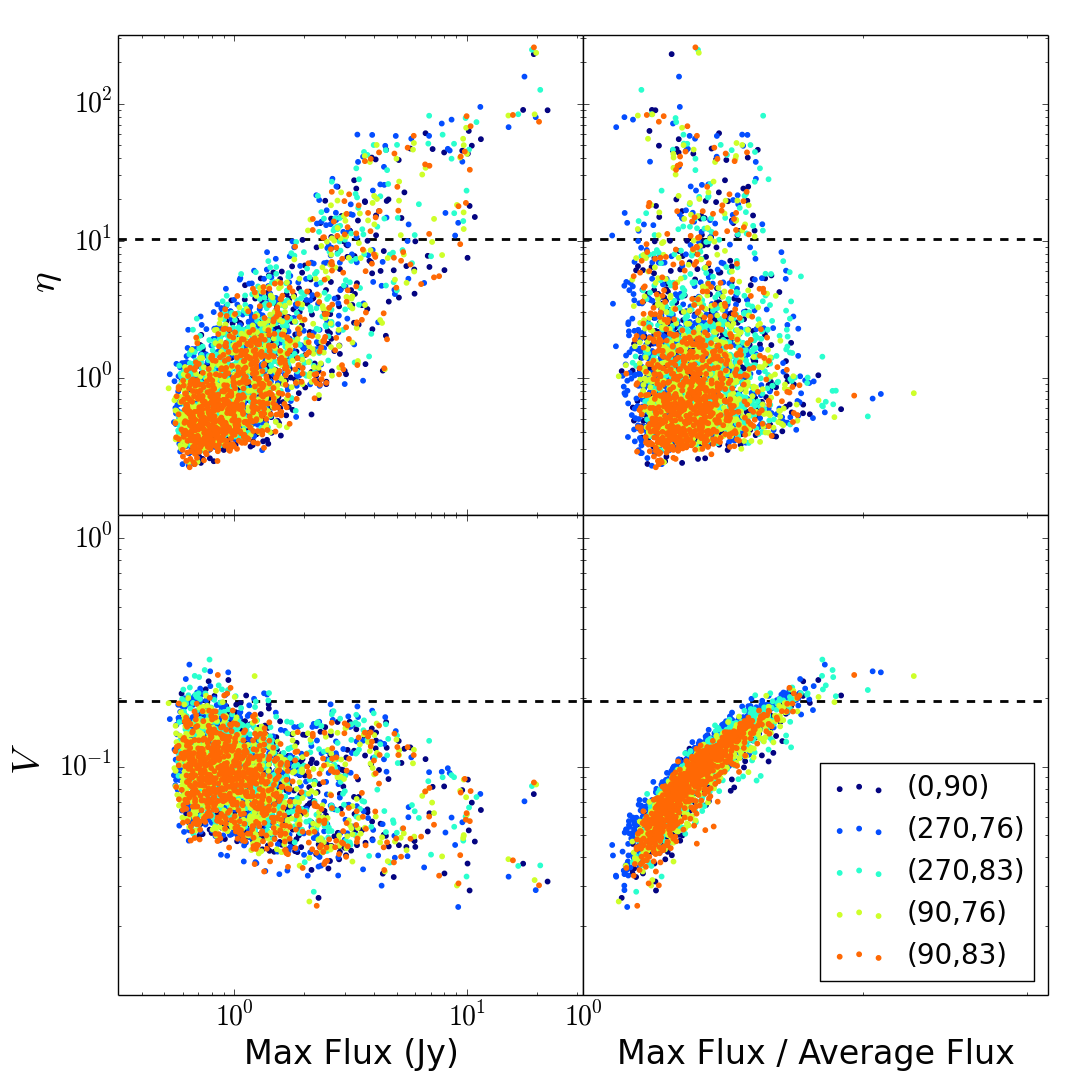}
\caption{Here we show the two variability parameters against the maximum flux density that each source attains and the ratio between this maximum flux density and the average flux density of the source. The colour scheme is as in Figure \ref{variPlt} and the dashed lines are the 2$\sigma$ thresholds used to identify variable sources.}
\label{variPlt2}
\end{figure*}

\subsection{Results}
\subsubsection{Blindly detected variability}

In Figure \ref{variPlt}, we show these variability parameters from the end of each {\sc TraP} run for each of the 5 unique azimuth-elevation pointing directions. As the timescales probed by each of the different pointing directions is roughly the same, it is expected that the typical source parameters for the different pointing directions will be in good agreement and this is clearly the case in Figure \ref{variPlt}. The $V$ distribution is well fitted with a Gaussian distribution, with a typical value of $V\sim0.1$, consistent with the typical flux density uncertainties measured in Section 2.8.  We note that the absolute $\eta$ values should be a factor $\sim$3 lower due to the correlated noise observed in these images (see Section 2.4); this does not affect the analysis in this section as we are only considering sources that are anomalous to the distribution. This factor will need to be considered when quantifying low level variability in future analysis. The $\eta$ distribution is clearly right-skewed with an excess of sources at higher values, suggesting that there may be variability in some of the source light curves. However, as the $V$ parameters of these sources are comparable to the rest of the population, this is unlikely (variable sources have anomalously high values for both variability parameters and are expected reside in the top right corner of this plot). By visual inspection, we note that many sources show variation on specific nights, pointing to a possible ionospheric origin or residual calibration issues. Further analysis is ongoing.

Using additional source parameters (particularly the maximum flux density that a source attains and the ratio between that maximum flux density and the average flux density of the source) can aid in understanding the population of sources and can be used to more clearly separate the variable sources from a stable population. In Figure \ref{variPlt2}, we plot these 4 parameters for each of the different pointing directions. We see a clear correlation between the maximum flux density and $\eta$ as expected (this is caused by the measurement accuracy of flux densities for bright sources, which does not take into account systematic uncertainties). Additionally, we observe a negative trend between the maximum flux densities and $V$, however there is a clear diversion from this trend at flux densities $>$2 Jy. This diversion is caused by a large number of images, coincident in time, having a systematic flux density scale offset (of order 10\%) from the rest of the images. This is likely caused by uncertainties in the primary beam model and is expected to be resolved in future work when we address the issues between different pointing directions. 

To avoid the region where the systematic uncertainties are dominating, we utilise a $2\sigma$ threshold on both the $\eta$ and $V$ parameters to identify variable sources. This corresponds to variable sources requiring $\eta \ge 10.6$ and $V \ge 0.20$, equivalent to a flux density variation in excess of 20\%. These thresholds will only identify sources which are significantly more variable than the typical population and does not address any intrinsic low level variability, such as the phenomenon of ``low frequency variability'' likely caused by refractive interstellar scintillation \citep[although this is not expected to occur in our dataset as it typically has timescales of $\gtrsim$1 year and is more prevalent at lower Galactic latitudes; e.g.][]{mitchell1994}. From the skewed $\eta$ distribution and visual inspection of a large number of light curves, we note that there is low-level variability on specific observing nights which may correspond to ionospheric activity. No significantly variable sources were identified via this method. 

\subsubsection{Known variables within the field}
Two sources within this field have observed variability, with timescales of 2 seconds at $\sim$150 MHz, which has been interpreted as Interplanetary Scintillation \citep[IPS; ][]{kaplan2015}. Using the zenith observations, we identify these two sources within our dataset and determine their variability parameters. PKS 2322-275 has $V=0.047$ and $\eta=6.2$ and PKS 2318-195 has $V=0.050$ and $\eta=2.0$. Both of these sources are well below the variability thresholds and their parameters are highly typical for the source population. \cite{kaplan2015} show that these sources have a typical variability timescale which is significantly shorter then the 30 second integration timescale used in this analysis. Therefore, any IPS events would be statistically averaged to the mean value in these images.

Additionally, there is one pulsar within the field, PSR 2327-20. This pulsar has a low DM and, hence, may undergo diffractive and refractive scintillation. In our median image, we note that this source is very faint, with a flux density of $\sim$0.07 Jy, and is unlikely to be detected in our images. We monitored the position of this pulsar to see if this pulsar scintillates above the detection threshold. PSR 2327-20 is not detected in any of the snapshot images.

\subsection{Future work}

Although this analysis has not identified any significant variability, we note that the variability analysis needs a significant amount of further work to be able to identify variability of sources corresponding to $\lesssim$20\% of their flux densities. For future analysis:

\begin{itemize}
    \item We intend to resolve remaining systematic primary beam uncertainties within the images. This will enable all the pointing directions to be processed at once, giving a much larger dataset for characterising the sources. Additionally, it will resolve the deviation in $V$ at flux densities in excess of 2 Jy, which will lead to an increase in sensitivity.
    \item The variability parameters for sources can change significantly as the number of data points in the light curve increase. For instance, if a source emits a single flare at early times this variability may not be apparent using the variability parameters from the final time step in the dataset. This means that we may be missing interesting variability on short timescales due to processing large numbers of images at once. {\sc TraP} records these variability parameters as a function of snapshot and, in future, we aim to develop methods to study the variability of sources as a function of time. 
\end{itemize}

\section{Conclusions}

From our analysis of $\sim$10,000 images, we note that the EoR0 field is remarkably stable at 182 MHz. There are no convincing transient candidates and all sources have flux density variations of $\lesssim$20\%. In future work, we will target remaining systematic flux density uncertainties to enable us explore low level variation within the field.

The transient surface densities obtained are more constraining than previous surveys by orders of magnitude in timescale, sensitivity and snapshot rates; although we note that this field is not sensitive to a Galactic population of transient sources due to being well off the Galactic plane (Galactic latitudes $\lesssim-65$ degrees). Despite expecting to observe transients comparable to the source observed by \cite{stewart2015}, we instead place a constraining limit on their surface densities and/or spectral indices. On the shortest timescale, predictions scaled from the observed populations suggested that this survey would identify a small number of FRBs. Again, there are no detections, which are consistent with suggestions of lower rates and flat spectral indices. 

To further pursue these elusive transients at low radio frequencies, we need to conduct similar surveys at a range of frequencies, whilst also increasing the sensitivities and surveyed area by an order of magnitude or more. Finally, there are a range of timescales that this survey does not explore, most notably the very short and $>$1 year where a range of transient sources are anticipated.

\section{Acknowledgements}

This scientific work makes use of the Murchison Radio-astronomy Observatory, operated by CSIRO. We acknowledge the Wajarri Yamatji people as the traditional owners of the Observatory site. Support for the MWA comes from the U.S. National Science Foundation (grants AST-0457585, PHY-0835713, CAREER-0847753, and AST-0908884), the Australian Research Council (LIEF grants LE0775621 and LE0882938), the U.S. Air Force Office of Scientific Research (grant FA9550-0510247), and the Centre for All-sky Astrophysics (an Australian Research Council Centre of Excellence funded by grant CE110001020). Support is also provided by the Smithsonian Astrophysical Observatory, the MIT School of Science, the Raman Research Institute, the Australian National University, and the Victoria University of Wellington (via grant MED-E1799 from the New Zealand Ministry of Economic Development and an IBM Shared University Research Grant). The Australian Federal government provides additional support via the Commonwealth Scientific and Industrial Research Organisation (CSIRO), National Collaborative Research Infrastructure Strategy, Education Investment Fund, and the Australia India Strategic Research Fund, and Astronomy Australia Limited, under contract to Curtin University. We acknowledge the iVEC Petabyte Data Store, the Initiative in Innovative Computing and the CUDA Center for Excellence sponsored by NVIDIA at Harvard University, and the International Centre for Radio Astronomy Research (ICRAR), a Joint Venture of Curtin University and The University of Western Australia, funded by the Western Australian State government.

This research was undertaken with the assistance of resources from the National Computational Infrastructure (NCI), which is supported by the Australian Government. This work was supported by the Flagship Allocation Scheme of the NCI National Facility at the ANU.

We gratefully acknowledge discussions with members of the LOFAR Transients Key Science Project when determining the optimal strategy for {\sc TraP} processing.

\appendix

\end{document}